Limits on Magnetized Quark-Nugget Dark Matter from Epi-sodic Natural Events

J. Pace VanDevender[1*], Aaron P. VanDevender [1], Peter Wilson [2], Benjamin F. Hammel [3] and Niall McGinley [4]

1     VanDevender Enterprises LLC, 7604 Lamplighter Lane NE, Albuquerque, NM 87109 USA; aaron@vandevender.com

2     School of Geography and Environmental Sciences, Ulster University, Cromore Road, Coleraine, Co. Londonderry, BT52 1SA Northern Ireland, UK; p.wilson@ulster.ac.uk

3     Department of Materials Science and Engineering, University of Pennsylvania, 3231 Walnut Street, Philadelphia, PA 19104, USA; bfhammel@seas.upenn.edu

4     Ardaturr, Churchill PO, Letterkenny, Co. Donegal, F928982, Ireland; niallmacfhionnghaile@gmail.com

*Correspondence: pace@vandevender.com

**Abstract:** A quark nugget is a hypothetical dark-matter candidate composed of approximately equal numbers of up, down, and strange quarks. Most models of quark nuggets do not include effects of their intrinsic magnetic field. However, Tatsumi used a mathematically tractable approximation of the Standard Model of Particle Physics and found that the cores of magnetar pulsars may be quark nuggets in a ferromagnetic liquid state with surface magnetic field $B_o$ = $10^{12\pm1}$ T. We have applied that result to quark-nugget dark matter. Previous work addressed the formation and aggregation of magnetized quark nuggets (MQNs) into a broad and magnetically stabilized mass distribution before they could decay and addressed their interaction with normal matter through their magnetopause, losing translational velocity while gaining rotational velocity and radiating electromagnetic energy. The two orders of magnitude uncertainty in Tatsumi's estimate for $B_o$ precludes the practical design of systematic experiments to detect MQNs through their predicted interaction with matter. In this paper, we examine episodic events consistent with a unique signature of MQNs. If they are indeed caused by MQNs, they constrain the most likely values of $B_o$ to $1.65 \times 10^{12}$ T +/- 21% and support the design of definitive tests of the MQN dark-matter hypothesis.

## 1. Introduction

The great majority of mass in the Universe is non-luminous material called dark matter [1]. Gravity from dark matter literally holds galaxies together [2]. The nature of dark matter has been studied for decades but remains one of the most puzzling mysteries in science [3]. Most dark-matter candidates are assumed to interact with normal matter only through gravity, but stronger interactions are consistent with requirements for dark matter if their effective interaction time is billions of years [3]; Magnetized Quark Nuggets (MQNs) are one such candidate for dark matter. Previously published work [4] shows the primordial origin of MQNs and their compatibility with requirements [5] of dark matter. Their origin in the very early Universe was in the magnetic aggregation of $\Lambda^0$ particles (consisting of one up, one down, and one strange quark) into a broad mass distribution of stable ferromagnetic MQNs before they could decay. After t ≈ 66 μs after the big bang, mean MQN mass is between ~ $10^{-6}$ kg and ~$10^4$ kg, depending on the surface magnetic field $B_o$. The corresponding mass distribution is sufficient for MQNs to meet the requirements of dark matter in the subsequent processes, including those that determine the Large Scale Structure (LSS) of the Universe and the Cosmic Microwave Background (CMB).

For the last four decades, searches for dark-matter candidates have focused on particles beyond the Standard Model of Particle Physics [6]. MQNs are composed of Standard Model quarks. However, mathematical techniques for applying the Model in the ~90-MeV-energy scale of MQNs rely on approximations. In addition, the key result that leads to MQNs requires a Quantum Chromo Dynamics (QCD) fine-structure constant $\alpha_c \approx 4$ at this energy scale; the value of $\alpha_c$ at this energy scale is not known.

To the extent that the calculations can be performed, MQNs are consistent with the Standard Model and do not require a new particle Beyond the Standard Model (BSM). Therefore, MQNs have been somewhat controversial as dark-matter candidates.

In the case of this paper, the controversy may be mitigated because we are not claiming discovery. The paper does find that the theory of MQNs is consistent with a reported observation. However, the event is very rare, is not reproducible, was not recorded by multiple observers, and cannot be quantitatively validated after the fact by anyone else. Consequently, the evidence does not meet today's standard for a discovery.

Quarks are components of many particles in the Standard Model of Particle Physics. Ensembles of strange, up, and down quarks (in approximately equal numbers) are called quark nuggets and are thought to be stable at sufficiently large masses [7] and qualify as candidates for dark matter. Quark nuggets are also called strangelets [8], nuclearites [9], Axion Quark Nuggets (AQNs) [10], slets [11], Macros [5], and MQNs [4]. A brief summary of four decades of research [4–35] on quark-nugget charge-to-mass ratio, formation, stability, and detection is provided for convenience and completeness as Appendix A: Quark-Nugget Research Summary.

Most models of quark nuggets do not include effects of their intrinsic magnetic field. However, Xu [26] has shown that the low electron density, permitted in stable quark nuggets, limits surface magnetic fields from ordinary electron ferromagnetism to ~2 x $10^7$ T. Tatsumi [27] examined ferromagnetism from a One Gluon Exchange interaction and concluded that the surface magnetic field could be sufficient to explain the ~$10^{12}$ T magnetic fields inferred for magnetar cores. Since the result depends on the currently unknown value of $α_c$ at the 90 MeV energy scale, the result needs to be confirmed with relevant observations and/or advances in QCD calculations.

Tatsumi's result has recently been applied to quark-nugget dark matter. By definition of ferromagnetic, the lowest energy state in Tatsumi's ferromagnetic liquid is with magnetic dipoles aligned. The individual quark nuggets are formed with baryon number A = 1, as are neutrons and protons, and may have spin and a corresponding surface magnetic field similar to that of protons and neutrons. However, unlike protons and neutrons, ferromagnetic dipoles of quark nuggets (MQNs) align upon aggregation and maintain the surface magnetic field. Their magnetic field at substantial distances is large because their aggregated size is large, not because their intrinsic magnetization (and corresponding surface magnetic field) is necessarily larger than that of other baryons.

Previous work [4] addressed the formation and aggregation of magnetized quark nuggets (MQNs) in the early Universe into a broad and magnetically stabilized mass distribution with baryon number $A$ between ~$10^3$ and $10^{37}$ before they could decay by the weak interaction; addressed the compatibility of MQNs with the requirements of dark matter; and addressed their interaction with normal matter through their magnetopause [28], while losing translational velocity, gaining rotational velocity, and radiating electromagnetic energy [36].

Electromagnetic energy accumulated in the Universe from MQNs is unfortunately not detectable because the plasma in most of the Universe is too low density to establish the magnetopause effect and the electromagnetic radiation from the rest of interstellar space is too low frequency to propagate through the solar-wind plasma and reach Earth. However, MQNs transiting through the plasma and gas around Earth spin up to MHz frequencies and should be detectable as they exit the magnetosphere [36]. Predicted event rates are strongly dependent on the MQN magnetic field $B_o$.

Uncertainty in Tatsumi's estimate of $B_o$ = $10^{12±1}$ T is too large to design a system to systematically look for MQNs. In this paper, we examine one type of episodic event consistent with a unique signature of MQNs, i.e., an MQN impacting Earth on a nearly tangential trajectory, penetrating the ground for a portion of its path, and emerging where it can be observed. We calculate what would be observed and compare the results with extant observations. Such episodic events are impossible to predict or reproduce and fall short of the standard for evidence of discovery in physics. Therefore, we are not asserting the discovery of MQNs but are examining consistency with MQNs and the resulting constraints on $B_o$. The results are useful for designing systematic tests of the MQN dark-matter hypothesis.

Without a reasonably small uncertainty in $B_o$, success in fielding systematic experiments is unlikely. We attempted such an experiment by instrumenting a 30 sq-km area of the Great Salt Lake in Utah, USA, and looked for acoustic signals from MQN impacts. No impacts were observed in 2200 hours of recording. Subsequent theory [4] explained the null result and showed that the predicted mass distribution of MQNs means that impacts are very rare. Even a planet-sized detector is marginal. Consequently, we have turned to observations of episodic, naturally occurring events to narrow the uncertainty in $B_o$.

Reference 4 shows the most important unknown in the theory of MQNs is the surface magnetic field parameter $B_o$, which quantifies the uncertainty in the distribution of MQN mass and the event rate. The mean of the surface magnetic field $<B_S>$ is related to $B_o$ in reference [4], through

$$<B_S> = \left( \frac{r_{QN}}{10^{18}(kg/m^3)} \right) \left( \frac{r_{DM\_T=100 Mev}}{1.6 \times 10^8 (kg/m^3)} \right) B_o \quad (1)$$

In Equation (1), $\rho_{QN}$ is the MQN mass density, and $\rho_{DM}$ is the density of dark matter at time $t \sim 65$ μs, when the temperature $T$ in the early Universe was ~ 100 MeV [37]. If better values of $\rho_{QN}$, $\rho_{DM}$, and $B_o$ are determined by observations, then a more accurate value of $<B_S>$ can be calculated with Equation (1).

In this paper, we show that a nearly tangential impact and transit through a chord of Earth by an MQN provides a unique signature: a magnetically levitated mass of greater than nuclear density that ionizes and excites the atmosphere for many minutes. We compare the results of analytic calculations and computer simulations of such an event to the observations on 6 August 1868, published by M. Fitzgerald [38] in the *Proceedings of the Royal Society*, the premier scientific publication at that time. The event's unusual characteristics are consistent with an extremely rare, nearly tangential, MQN impact. Similar events have been reported elsewhere [39], yet Fitzgerald's report is the best documented and only scientifically published event we have found, making it the most suitable for comparison with theory.

As noted above, Tatsumi [27] estimated that magnetar cores have $B_o = 10^{12 \pm 1}$ T. The results reported in this paper and reference 4 constrain the most likely values of $B_o$ to $1.65 \times 10^{12}$ T +/- 21% and will permit the design of a systematic experiment to test the MQN hypothesis.

## 2. Materials and Methods

This study used:
1. Analytic methods presented in detail in the results section.
2. Computational simulations coupling the Rotating Magnetic Machinery module and the Nonlinear Plasticity Solid Mechanics module of the 3D, finite-element, COMSOL Multiphysics code [40]. Details are included in Appendix B: COMSOL Simulation of Rotating Magnetized Sphere Interaction with Plastically Deformable Conductor.
3. Original field work at the location reported by Fitzgerald [38] in County Donegal, Ireland, is documented in Appendix C: Field Investigation of Fitzgerald's Report to Royal Society. The GPS locations are included to facilitate replication, subject to acquiring permission from the property owners listed in Acknowledgements. Radiocarbon dating was conducted by Beta Analytic Inc. 4985 SW 74th Court, Miami, Florida 33155, USA.

## 3. Results

*3.1. Nearly Tangential Impact and Transit of MQNs through Earth*

Figure 1 illustrates three MQN impacts on an idealized portion of a spherical Earth.

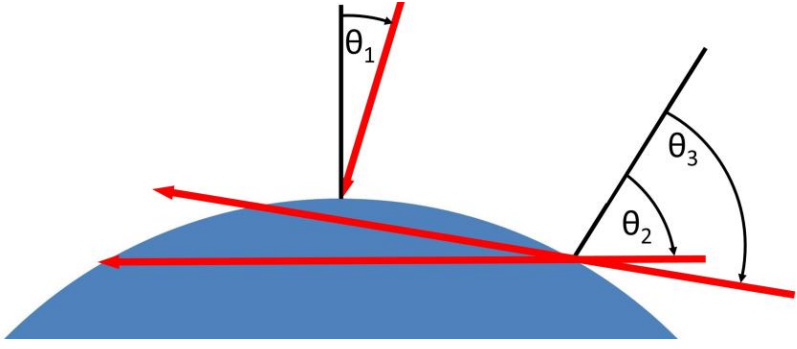

**Figure 1.** Three MQN trajectories are shown as red arrows, along with their angle of impact $\theta$ with respect to the normal surface of an idealized, not to scale, Earth (blue). The trajectory with impact angel $\theta_1 \ll 90°$ is a more common radial impact. Nearly tangential trajectories that transit through Earth are represented by trajectories with impact angles between $\theta_2$ and $\theta_3$. MQNs on a $\theta_2$ trajectory emerge from Earth with negligible velocity after transiting a distance $x_{max}$, the maximum range of an MQN. MQNs on a $\theta_3$ trajectory emerge from Earth with considerable velocity.

Only massive MQNs have enough momentum to penetrate significant distances through water or rock. Therefore, we focus on MQNs between $10^5$ kg (maximum mass associated with $B_o \sim 1.3 \times 10^{12}$ T) and $10^{10}$ kg (maximum mass associated with $B_o \sim 3 \times 10^{12}$ T), and use these extremes to illustrate each calculation.

*3.2. Slowing Down in Passage through a Portion of Earth*

Hypervelocity MQNs ionize surrounding matter through a shock wave and interact with that matter through a magnetopause in the same way that Earth interacts with the solar wind through its magnetopause. The equations governing the interaction are derived in reference [28] and are summarized in Equations (2) through (6) for convenience.

The force equation for a high-velocity body with instantaneous radius $r_m$, and velocity $v$, moving through a fluid of density $\rho_p$ with a drag coefficient $K \sim 1$, is

$$F_e \approx K \pi r_m^2 \rho_p v^2 \qquad (2)$$

The geometric radius $r_{QN}$ of the MQN of mass $m$ and mass density $\rho_{QN}$ is

$$r_{QN} = \left( \frac{3m}{4\pi \rho_{QN}} \right)^{\frac{1}{3}} \qquad (3)$$

However, MQNs have a velocity-dependent interaction radius [28] equal to the radius of their magnetopause

$$r_m \approx \left( \frac{2B_o^2}{\mu_0 K \rho_p v^2} \right)^{\frac{1}{6}} r_{QN} \qquad (4)$$

MQNs with mass $10^5$ kg have $r_{QN} \sim 4 \times 10^{-5}$ m. For $B_o \sim 1.3 \times 10^{12}$ T and $v = 250$ km/s, an MQN passing through water of density 1000 kg/m$^3$ has the magnetopause radius $r_m \sim 0.025$ m. The corresponding values for mass $m = 10^9$ kg with $B_o \sim 3 \times 10^{12}$ T are $r_{QN} \sim 9 \times 10^{-4}$ m and $r_m \sim 0.71$ m. Although their nuclear density makes these massive MQNs physically small, their large magnetic fields and high velocities make their interaction radius and cross section very large, even in solid-density matter.

The interaction radius of an MQN varies as velocity $v^{-1/3}$ in Equation (4). Including that velocity dependence in the calculation with initial velocity $v_o$ gives velocity $v$ as a function of distance $x$:

$$v = v_o \left(1 - \frac{x}{x_{max}}\right)^{\frac{3}{2}} \qquad (5)$$

in which $x_{max}$ is the stopping distance for an MQN:

$$x_{max} = \left(\frac{3m}{2\pi r_{QN}^2}\right)\left(\frac{\mu_o v_o^2}{2K^2 \rho_p^2 B_o^2}\right)^{\frac{1}{3}} \qquad (6)$$

MQNs with mass $10^5$ kg, $B_o \sim 1.3 \times 10^{12}$ T, and velocity $v_o \sim 250$ km/s have $x_{max} \sim 241$ km passing through water. The corresponding value for mass of $10^9$ kg with $B_o \sim 3 \times 10^{12}$ T is $x_{max} \sim 3000$ km.

From the geometry in Figure 1, the angle of incidence $\theta_2$ for trajectories that emerge from Earth with negligible velocity is given by

$$\theta_2 = \cos^{-1}\left(\frac{x_{max}}{2r_{Earth}}\right) \qquad (7)$$

in which $r_{Earth}$ is the radius of the Earth, or $\sim 6.38 \times 10^6$ m. For $\pi/2 > \theta > \theta_2$, MQNs emerge from Earth with velocity

$$v_{exit} = v_o \left(1 - \frac{x_{exit}}{x_{max}}\right)^{\frac{3}{2}} \text{ for } x_{exit} = 2r_{Earth} \cos(\theta) \qquad (8)$$

Integrating Equation (5) gives the transit time $t_{exit}$

$$t_{exit} = \left(\frac{2x_{max}}{v_o}\right)\left(\left(\frac{x_{max}}{x_{max} - x_{exit}}\right)^{\frac{1}{2}} - 1\right) \qquad (9)$$

Transit time $t_{exit}$ is strongly dependent on ($x_{max} - x_{exit}$) and is infinite at $x_{max} = x_{exit}$. A $10^5$ kg MQN with $B_o = 1.3 \times 10^{12}$ T, initial velocity $v_o = 250$ km/s, and incidence angle $\theta_2 = 88.91817°$ penetrates a distance $x_{exit} = 240.09$ km of water and emerges in $t_{exit} \sim 54$ s with $v_{exit} = 10$ m/s. If the incidence angle $\theta_2 = 88.92278°$, then $x_{exit} = 239.89$ km, transit time $t_{exit} \sim 24$ s, and $v_{exit} = 100$ m/s.

During the transit, the MQN is falling towards the center of Earth with acceleration $g = 9.8$ m/s², which is not considered in Equations (7) through (9). The deviation from the straight-line approximation is $\delta r \sim \frac{1}{2}gt_{exit}^2$ and the corresponding fractional error in path length introduced by neglecting gravity is

$$\delta \sim \frac{\delta r}{\sqrt{x_{exit}^2 + (\delta r)^2}} \qquad (10)$$

For the example in the previous paragraph, fractional error in path length because of gravity is $\delta \sim 0.06$ for $v_{exit} = 10$ m/s and is $\delta \sim 0.012$ for $v_{exit} \sim 100$ m/s. In general, fractional error decreases with increasing $v_{exit}$, decreasing $B_o$, decreasing MQN mass, and increasing mass density of material transited (granite with $\rho_p = 2{,}300$ kg/m³ or water $\rho_p = 1000$ kg/m³).

If the MQN slows to <<10 m/s, gravity dominates its motion, and it does not emerge from the ground or water. Fast objects are difficult to perceive, especially if an observer is not primed to expect an event. A human observer requires ~ 0.25 sec to perceive an object as a thing [41]. If the object is about 1 m in diameter and moving at > 100 m/s, the object will have moved > 25 m before cognitive acquisition and may be out of range before the observer can process the image well enough to be confident of what was seen. Therefore, we suggest that only MQNs emerging with 10 m/s ≤ $v_{exit}$ ≤ 100 m/s are likely to be reported by human observers.

Rearranging Equation (8) gives the incidence angle $\theta_{vexit}$ for a given $v_{exit}$:

$$\theta_{vexit} = \cos^{-1}\left(\left(\frac{x_{max}}{2r_{Earth}}\right)\left(1-\left(\frac{v_{exit}}{v_o}\right)^{\frac{2}{3}}\right)\right) \quad (11)$$

For $v_{exit}$ = 10 m/s and 100 m/s, the corresponding impact angles are, respectively, $\theta_{10}$ and $\theta_{100}$, which will be used in estimating the event rates for directly observable MQN events.

*3.3. Estimated Event Rates*

The number of events per year on Earth is estimated as follows:
1. Earth is moving about the galactic center, in the direction of the star Vega, and through the dark-matter halo with a velocity of ~230 km/s [42]. Therefore, dark matter streams into the Earth frame of reference with mean streaming velocity ~230 km/s.
2. Dark matter in the halo also has a nearly Maxwellian velocity distribution with mean velocity of ~230 km/s, so the ratio of streaming velocity to Maxwellian velocity is approximately 1 [42].
3. Approximating the velocity of dark matter streaming from the direction of Vega as ~230 km/s, we calculate the cross section $A_{10\text{-}100}$ for transiting a chord through Earth and emerging with velocity between $v_{10}$ = 10 m/s and $v_{100}$ = 100 m/s:

$$A_{10-100} = 2\pi r_{Earth}^2 \left(\sin\theta_{100} - \sin\theta_{10}\right) \quad (12)$$

More generally, the cross section for MQNs impacting between $\theta_{min}$ and $\theta_{max}$ is

$$A = 2\pi r_{Earth}^2 \left(\sin\theta_{max} - \sin\theta_{min}\right) \quad (13)$$

4. MQNs can have masses between $10^{-23}$ kg and $10^{10}$ kg [4]. We approximate such a large range by 1) associating the flux of all MQNs that have mass between $10^i$ kg and $10^{i+1}$ kg with a representative mass $10^{i+0.5}$ kg (which we call the representative decadal mass) for $-23 \leq i \leq 10$; 2) calculating the behavior of each decadal-mass MQN; 3) assuming all the MQNs in that decadal range behave the same way. The associated number flux is called the decadal flux $F_{m\_decade}$ (number N/y/m$^2$/sr) and was computed [4] as a function of $B_o$ from simulations of the aggregation of quark nuggets from their formation in the early Universe and evolution to the present era.
5. For $A_{10\text{-}100\_m\_decade}$, defined as the $A_{10\text{-}100}$ appropriate to a decadal mass $m$, the number of events per year per steradian for MQNs streaming from the direction of Vega and emerging with velocity between 10 and 100 m/s is $F_{m\_decade} A_{10\text{-}100\_m\_decade}$, summed over all decadal masses $m$.
6. For random velocity, approximately equal to streaming velocity, reference [36] shows that 5.56 sr is the effective solid angle that generalizes the streaming result to include MQNs from all directions.

7. Therefore, the total number of events per year somewhere on Earth with $v_{exit}$ between 10 and 100 m/s is

$$N = 5.56 \sum_{m\_decade} (F_{m\_decade} A_{10-100\_m\_decade})  \tag{14}$$

In Section 3.6, we will show that MQNs transiting Earth along a chord spin up to MHz frequencies and emit substantial radio frequency (RF) power. If and only if the RF is not absorbed by the surrounding plasma, these MQNs can be detected by their RF emissions propagating around Earth in the waveguide between the ground and the ionosphere. Their cross section is given by Equation (13) with $\theta_{max} = \pi/2$ and $\theta_{min} = \theta_2$ from Equation (7).

Figure 2 shows the estimated number of events per year somewhere on Earth as a function of $B_o$. Two modes of transit (through granite or water) and both potential modes of detection (human or radiofrequency) are considered.

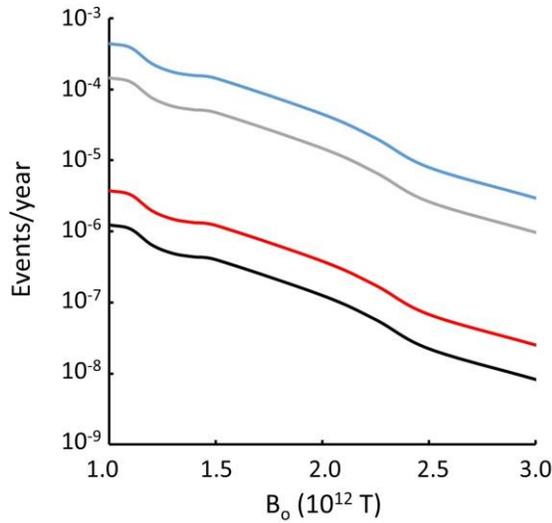

**Figure 2.** Estimated number of events per year somewhere on Earth as a function of $B_o$ for MQNs with $10^5$ kg $\leq m \leq 10^{10}$ kg. The four solid-line curves correspond to event rates based on interstellar dark-matter density [3,42] of ~7 × $10^{-22}$ kg/m³: MQNs transiting through water and emerging with any velocity $v_{exit}$ (blue); MQNs transiting through water and emerging with velocity 10 m/s $\leq v_{exit} \leq$ 100 m/s (gray); MQNs transiting through granite and emerging with any velocity $v_{exit}$ (red); MQNs transiting through granite and emerging with velocity 10 m/s $\leq v_{exit} \leq$ 100 m/s (black).

Figure 2 shows that detection on or near water is much more likely than detection deep within continents and shows the event rate is strongly dependent on parameter $B_o$. Unless the RF is absorbed by the surrounding plasma, the detection of MQNs by RF instruments (sensitive to any velocity) is much more likely than the detection of MQNs at 10 to 100 m/s, observable by human observers; however, records of human observations span centuries. Even one reliable report of an MQN event with the characteristics of a nearly tangential transit would suggest low values of $B_o$ and a mechanism that enhances the density of dark matter inside the solar system compared to that of interstellar space, as briefly described in Section 4.5.

*3.4. Rotation at Megahertz Frequencies*

MQNs interact with matter through its magnetopause. The shape of the magnetopause depends on the angle between the MQN velocity and the MQN magnetic moment, as illustrated in Figure 3.

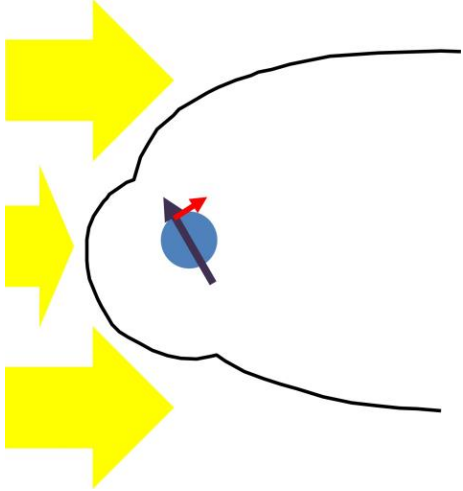

**Figure 3.** Cross sectional view of the magnetopause is shown (black line) between an MQN (blue circle) with magnetic moment (purple vector) at an angle of 60° to the velocity of the plasma (yellow arrows) flowing into the rest frame of the MQN. The plasma flow produces a net force (red arrow) centered at the top of the MQN and a corresponding torque vector into the page. The magnetopause is the locus of points at which the plasma pressure (on the left in Figure 3) is balanced by the magnetic pressure of the compressed magnetic field on the right. The complex shape of the magnetopause and resulting torque have been computed by Papagiannis [43] for Earth, illustrated in Figure 3, and extended to the case of MQNs. The effect can be understood by considering that the mean distance between the magnetopause and the MQN on the top half of Figure 3 is less than on the bottom half, which means that the magnetic field is compressed more on the top than on the bottom. Since force is transmitted by the compressed magnetic field, the net force is a push on the top, as shown by the red arrow.

Time-dependent asymmetry of the magnetopause produces a velocity-dependent and angle-dependent torque on the MQN and causes the MQN to oscillate initially about an equilibrium [36], as shown in Figure 4. Since the quark nugget slows down as it passes through ionized matter, the decreasing forward velocity reduces the torque with time, so the time-averaged torque in one half-cycle is greater than the opposing time-averaged torque in the next half-cycle. The amplitude of the oscillation necessarily grows. Once the angular momentum is sufficient to give continuous rotation, the net torque continually accelerates the angular motion to produce a rapidly rotating quark nugget. As shown in Figure 4, MHz frequencies are quickly achieved, even with a $10^6$ kg quark nugget moving through granite (2300 kg/m$^3$ density matter) by the time $v$ has slowed to 220 km/s.

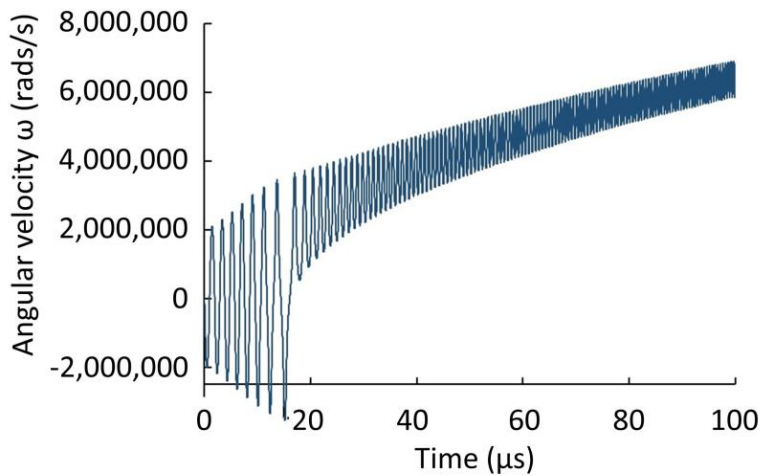

**Figure 4.** Estimated angular velocity in the first $10^{-4}$ s for $10^6$ kg quark nugget with velocity $v = 220$ km/s, initial angle $\chi = 0.61$ rad, initial angular velocity $\omega = 0$, and passing through matter with mass density of 2300 kg/m³. Note the initial oscillation about 0 until a full rotation occurs, after which the angular velocity increases rapidly.

As developed in reference [36] and summarized in Equations (15) through (19) for convenience, hypervelocity MQNs transiting through matter experience a torque that causes them to rotate with a frequency that depends on $B_o$, MQN mass $m$, MQN velocity $v$, and density $\rho_p$ of the surrounding material. Rotating magnetic dipoles radiate at power $P$, where

$$P = \frac{Z_o}{12\pi}\left(\frac{\omega}{c}\right)^4 m_m^2 \qquad (15)$$

in SI units, with $Z_o = 377$ Ω, $\omega$ = angular frequency, and $c$ = the speed of light in vacuum. Magnetic dipole moment $m_m = 4\pi B_o r_{QN}^3/\mu_o$.

Assuming the energy loss per cycle from electromagnetic radiation to the surrounding plasma is just balanced by the energy gain per cycle from the torque $T$,

$$\int_0^{\frac{2\pi}{\omega}} T dt = \frac{2\pi P}{\omega^2} \qquad (16)$$

in which

$$T = C_2 \rho_p^{0.5} v B_o r_{QN}^3 F_\chi$$
$$F_\chi = MIN(ABS(\tan \chi), ABS(\cot \chi))\frac{\tan \chi}{ABS(\tan \chi)} \qquad (17)$$

The constant $C_2 = 1400$ with units of N s kg$^{-0.5}$ m$^{-1.5}$ T$^{-1}$, and the angle of rotation $\chi$ is the angle between the velocity of the incoming plasma and the magnetic moment.

The rate of change of angular velocity $\omega$ for MQN of mass $m$, with moment of inertia $I_{mom} = 0.4\, m\, r_{QN}^2$, and experiencing torque $T$, is

$$\frac{d\omega}{dt} = \frac{T}{I_{mom}} \qquad (18)$$

Combining Equations (15) through (18) gives

$$\int_0^{\frac{2\pi}{\omega}} \frac{v(t) F_\chi(\chi(\omega t))}{\omega^2} dt = \frac{5.54 \times 10^{-22} B_o r_{QN}^3}{\rho_x^{0.5}} \qquad (19)$$

which is solved numerically for the equilibrium angular velocity $\omega$ or frequency $f = \omega/(2\pi)$.

After emerging from dense matter, MQN rotational frequency decreases as rotational energy is radiated away with power $P$. Since $P$ varies as $\omega^4$ and the rotational energy varies as $\omega^2$, the frequency as a function of time is not exponential. Solving for $\omega(t)$ gives

$$C_{RF} = \frac{1.08 \times 10^{43}}{B_o^2 m^{\frac{1}{3}}}$$

$$\omega(t) = \omega_0 \sqrt{\frac{C_{RF}}{\omega_0^2 t + C_{RF}}}$$

(20)

Representative results for slowing down during transit through water and granite and for parameters of greatest interest are shown in Appendix D: Tables of MQN Interactions with Water and Granite.

The tables show RF frequencies at emergence from water or granite are weakly dependent on $B_o$ but strongly dependent on mass and range from 7 MHz to 0.3 MHz for mass $m$ between $10^5$ kg and $10^{10}$ kg, respectively. Rotational energy ranges from ~0.1 MJ to ~1000 MJ and equals ~$10^{-11}$ times the translational energy at impact. RF power emission at emergence ranges from ~4 GW to 22 TW.

The tables also show that the RF power, calculated with Equations (14) and (19) at $t = 1200$ seconds after emergence, ranges from ~6 MW to ~200 GW for the most massive MQNs with $B_o$ between $1.3 \times 10^{12}$ T and $3.0 \times 10^{12}$ T, respectively.

Event rates in Figure 2 assume MQNs have the same flux as interstellar dark matter. However, the MQN flux may exceed that of interstellar space. Some MQNs passing through a portion of the solar photosphere are slowed to less than escape velocity from the solar system. Some of these are subsequently deflected by the net gravity of the planets so that they cannot return to the Sun and be absorbed. They accumulate. Therefore, this aerocapture process can enhance the local flux of MQNs and make Figure 2 a worst-case scenario. Enhancement factor, multiplied by observation time, would have to be >> 1000 for MQNs with nearly tangential trajectories to be observed.

Since RF detection occurs in real time, Figure 2 shows that the predicted event rate is too low to use RF to observe MQNs with nearly tangential trajectories, even if the RF is not absorbed by the plasma surrounding the MQN. Therefore, RF detection is best done in space [36] where the cross section is much larger and RF cutoff can be avoided.

In contrast, direct observations by human observers can cover centuries and may be recorded for us to analyze. The next section explores the observables in such an event to compare with observations published in the *Proceedings of the Royal Society* regarding an event on 6 August 1868 [38].

*3.5. Simulations of a Rotating MQN with Plastically Deformable Conducting Witness Plate*

The force between a rotating magnetized sphere and a plastically deformable conducting material was simulated by coupling the Rotating Magnetic Machinery module and the Nonlinear Plasticity Solid Mechanics module of the 3D, finite-element, COMSOL Multiphysics code [40]. The geometry is illustrated in Figure 5.

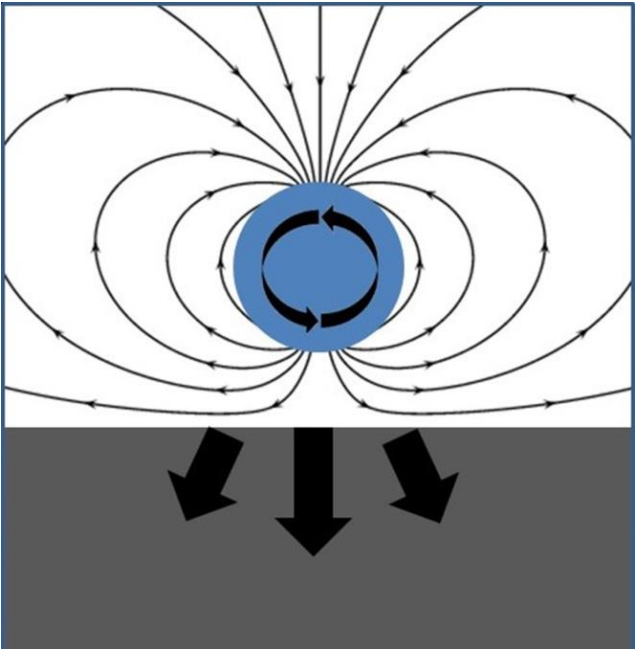

**Figure 5.** Geometry of simulation of rotating magnetized sphere above a highly conducting material. Magnetized, rotating sphere (blue) is shown above conducting material (gray). Arrows inside sphere indicate rotation and arrows in conducting material indicate force on the material. Arrows in air (white) indicate magnetic field lines at one moment in time. The axis of rotation is the *y*-axis, out of the plane of the figure. The magnetic axis of the magnetized sphere is initially in the *x*-direction and remains in the *xz*-plane.

Details of the simulation are provided in Appendix B: COMSOL simulation of rotating magnetized sphere interaction with plastically deformable conductor.

A small-scale experiment validated the COMSOL force calculation. A 3 mm or 6 mm thick copper plate was placed below a rotating spherical magnet and suspended with a calibrated spring to measure the force on the plate as a function of 1) separation between the center of the sphere and the front of the plate and 2) rotational frequency of the sphere. The measured force agreed with the force computed by the COMSOL simulation to within 10%. The agreement validates the computational method with the rotating coordinate system in the COMSOL module.

The computational mesh cannot resolve the microscopic diameter of an MQN. Therefore, we approximate the MQN with a 0.1 m radius, magnetized sphere with surface magnetic induction $B$ = 17,000 T, which corresponds to an MQN with mass $m = 7 \times 10^7$ kg and $B_o = 1 \times 10^{12}$ T.

Simulations with different values of electrical conductivity $\sigma$ and frequencies $f$ showed that the dynamics of the problem scales with the electromagnetic skin depth $\lambda$:

$$\lambda = \frac{1}{\sqrt{\pi \mu_0 \sigma f}} \tag{21}$$

Simulations converged best with low frequency. Consequently, we used frequency $f$ = 10 Hz and varied the conductivity $\sigma$ to explore the time-averaged force as a function of $\lambda$. The scaling with $\lambda$ let us apply the results to higher frequencies and more realistic values of $\sigma$. Distance between the center of the sphere at $z$ = 0 and the surface of the simulated peat was $z_p$ = -0.3 m. Results are shown in Figure 6.

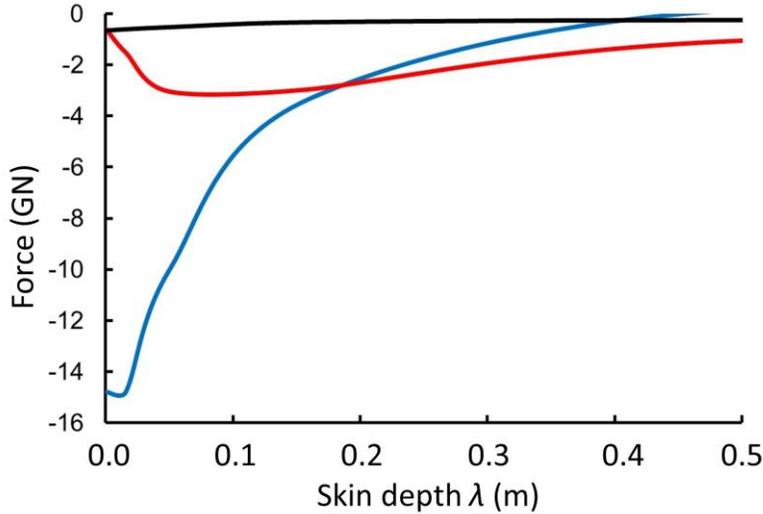

**Figure 6.** Components of the time-averaged force between the simulated quark nugget and conducting slab as a function of the electromagnetic skin depth $\lambda$. The negative (<0) force $F_z$ (blue) opposes gravity and levitates the rotating magnetized sphere for $\lambda < 0.5$ m, with the most negative value for $\lambda < 0.03$ m. The force generated by the magnetic field traveling through the deformable conductor in the $x$-direction, as the magnetized sphere rotates about the $y$-axis, generates a propulsive force $F_x$ (red). $F_x$ is much less than $F_z$ for small skin depths. The much smaller $F_y$ (black) illustrates ±5% error in the calculation, since symmetry requires $F_y = 0$.

The levitating force is strongly dependent on the skin depth and decreases rapidly with increasing skin depth. A skin depth of 0.05 m, corresponding to $\sigma = 10^7$ S/m and $f = 10$ Hz for $\sigma f = 10^8$ SHz/m, was chosen for the simulation of plastic deformation.

The radius of the magnetized sphere was set at 0.1 m and its magnetic induction field was set at 2085 T, corresponding to an MQN with mass $\sim 9 \times 10^7$ kg and $B_o = 1 \times 10^{12}$ T. The radius of the rotating volume in the COMSOL mesh was set at $r = 0.2$, and the front surface of the $4 \times 4 \times 2$ m deformable conductor was at $r = 0.3$ m.

For a time-averaged force of $10^7$ N in the $z$-direction (the direction opposing gravity), the maximum magnetic induction in the peat is 18.5 T and the maximum induced current density is $3 \times 10^8$ A/m². The time-averaged forces were $-1.1 \times 10^7$ N, $-0.3 \times 10^7$ N, and $-0.05 \times 10^7$ N in the $z$-, $x$-, and $y$-directions, respectively.

Deformation of the material as a function of time from the integrated simulation is shown in Figure 7 as contour plots for four times and two orthogonal directions.

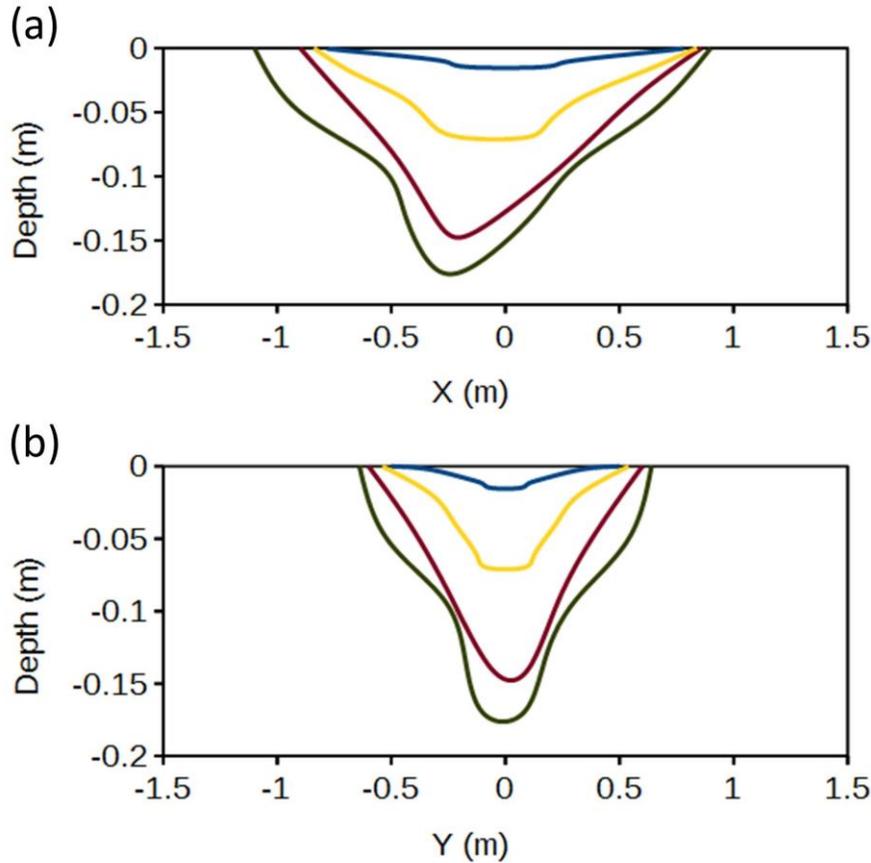

**Figure 7.** Contours of the hole formed by the rotating the magnetic field of the magnetized sphere in (a) the *x*-direction and (b) the *y*-direction for times 2.5 ms (blue), 10 ms (gold), 20 ms (red), and 30 ms (black). The magnetic field sweeps through plastically deformable conducting material, and the displacement of the bottom of the hole is approximately -0.25 m in the *x*-direction. The same contours for the *y*-direction, which is along the axis of rotation, show the deformation is symmetric about *y* = 0, as expected. In both cases, the vertical axis has a different scale from the horizontal axis.

A rotating magnetic dipole is equivalent to two oscillating current loops oriented at $90^0$ and with currents $90^0$ out of phase. The linear superpositions of the two magnetic fields induce currents that are almost independent of orientation but produce a net force in the direction (x) perpendicular to the axis of rotation (y). That explains the shape of the deformations in Figures 7a, b.

The simulation has many limitations. The MQN cannot be realistically resolved, so the gradient in the *B* field is not exact. The rotating sphere cannot move downward as material is displaced, so the deformation stops when the applied stress reaches equilibrium with the strain specified in the stress–strain curve. The yield strength of the material is a constant, independent of the degree of compression and of flow of liquid driven by the **J** × **B** force in the material, so the deformation rate is only qualitative. Consequently, these simulations provide only semi-quantitative results to compare with observations.

Despite these limitations, these simulations clearly show that a rapidly rotating magnetized sphere with sufficient mass and sufficient magnetic dipole field, such as a massive MQN, will create a hole in a plastically deformable conducting material by currents induced in its interior. They also show the sphere will experience a force that moves it through the material to create a trench in conducting, deformable material.

In addition, the results in Figure 6 clearly indicate that the levitating force decreases rapidly with increasing skin depth, which varies inversely with the square root of frequency. Therefore, the MQN height above conducting material decreases as the rotational energy is depleted by RF emissions and by the resistive dissipation of current induced in the peat.

The RF power emissions are predicted to be megawatts to many gigawatts and are certainly sufficient to ionize and excite the surrounding air. Thermal and magnetohydrodynamic motion and mixing of the air around an MQN complicates simple estimates of the shape of the luminosity; however, Equation (15) shows that the radiated power varies as the fourth power of the frequency, so the local electric field varies as the square of the frequency. Therefore, the ionization and excitation of air will diminish faster than the height above the peat.

These simulations assume induced current flows in the conducting medium. If the surface electric field, from the rate of change of magnetic induction, is sufficient to form a plasma at the air–ground interface, then magnetic pressure will deform the material as if $\lambda \sim 0$.

*3.6. Comparison with M. Fitzgerald's Report to the Royal Society*

M. Fitzgerald's report [37] to the Royal Society describes an event that occurred on 6 August 1868. The reported observations are consistent with the characteristics of a nearly tangential MQN impact as developed in this paper: a luminous orb with clear skies overhead, persisting for much longer than weather-related ball lightning [44], moving slowly across and into a plastically deformable conducting medium (peatland), decreasing in diameter with time, and creating trenches in the ground. Since it is difficult to obtain copies of proceedings more than a century old, Fitzgerald's description is quoted from the published report in Appendix C: Field Investigation of M. Fitzgerald's Report to Royal Society. In summary, a ~0.60 m diameter, light-emitting orb was observed travelling at ~1 m/s over and into a peat bog in County Donegal, Ireland, for ~20 minutes. During that time, its diameter decreased to ~0.08 m; it displaced over $10^5$ kg of water-saturated peat; it produced approximately 1 m wide trenches in the peat.

Although we initially dismissed his report as incredible, we eventually realized that peat only grows a few centimeters per century [45], so the features he described must still be visible if the reported event actually occurred. His report was sufficiently detailed to enable an investigation, which we conducted from 2004 through 2006.

Fitzgerald's reported features and our corresponding findings are summarized by the following:

- An approximately 6.4 m square hole described by Fitzgerald on the course from the crown of the ridge to the south of Meenawilligan, towards the town of Church Hill. We found a 6.4 m square hole 0.7 m deep along that course.

- An approximately 180 m distance reported to the next deformation. We found the deformation had been partially destroyed by draining of the field for sheep grazing. If this deformation were still the reported 100 m length, the southern end would be 175 ± 2 m from the hole.

- An approximately 100 m long, 1.2 m deep, and 1 m wide trench. As stated above, this deformation has been truncated by the owner having drained the field. The remaining trench is currently 63 ± 1 m long, 0.2 ± 0.05 m deep (soft to 0.8 ± 0.05 m), and 1.2 ± 0.1 m wide. Carbon dating of peat inside and outside the trench confirms a disturbance occurred, consistent with the report.

- Unspecified distance to the third excavation. We found the distance to be 5 ± 0.3 m.

- Curved trench formed when the stream bank was "torn away" for 25 m and dumped into the stream. We found the remaining curved trench to be 25 ± 1 m long and 1.4 ± 0.1 m deep. The 1863 Ordnance Survey map does not show the stream diversion that Fitzgerald reported as occurring on August 6, 1868. Therefore, the event happened after 1863. Fitzgerald's submission to the Royal Society is dated March 20, 1878, so the event occurred before 1868. Therefore, the event is independently dated between 1863 and 1878.

- Cave in the stream bank directly opposite the end of the "torn away" bank. We found the cave at that position. It is currently $0.45 \pm 0.08$ m wide, $0.3 \pm 0.06$ m high, and $0.5 \pm 0.1$ m deep. However, its proximity to the water line raises the possibility that its origin was flowing water and not the event Fitzgerald reports.

The extant features support Fitzgerald's account. This part of Ireland was, and still is, sparsely populated and did not have local newspapers that might have recorded an unusual sound or tremor. Therefore, no contemporaneous and independent eyewitness report confirms his account.

The vegetation is dominated by members of the heath (Ericaceae) and sedge (Cyperaceae) families. Surface features would become obscured to some degree by the growth of vegetation and debris carried in by wind, reducing the time a depression could be observed from a distance. A nearby crater, attributed [47] to a vertical impact of a 10 kg MQN in 1985, has changed little in the past 35 years, which is consistent with the current state of the features Fitzgerald attributed to the 1868 event.

The size of the trenches and the yield strength of peat gives a downward force of $>10^7$ N, which implies a rotating, magnetically levitated mass of $\sim 10^6$ kg. That mass and the volume of the $\sim$0.08 m diameter luminosity implies a mass density $>10^9$ kg/m$^3$, which is inconsistent with normal matter. Its levitation implies an extremely large magnetic dipole rotating at >1 MHz to levitate the large mass.

The repulsive force of electromagnetic induction by an MQN, as derived in this paper, are consistent with the levitation of an MQN core and the displacement of the peat in its path. The yield strength of peat was measured and found to be 530 kN/m$^2$ ± 23%, which is the value used in the simulations of plastic deformation above. Electrical conductivity $\sigma$ within 0.2 m of the surface was measured to be 22 mS/m ± 30%, which is consistent with published values for peatlands [46], as follows: 25 mS/m near the surface and ~380 mS/m at up to 2 m depth. Table A2 in Appendix D (Tables of MQN Interactions with Water and Granite) gives 3 to 9 MHz for the frequency of a massive MQN when it first emerges from the ground. The corresponding skin depth $\lambda \sim 1.1$ to 2.0 m for $\sigma = 22$ mS/m and $\lambda \sim 0.27$ to 0.47 m for $\sigma = 385$ mS/m. As shown in Figure 6, these values of $\lambda$ are too large to produce the reported levitation and deformation if the induced current is flowing through the peat, as assumed in the simulations.

However, the large electric field from the rate of change of the magnetic induction should cause the surface to flash over and form a plasma on top of the peat, similar to how the air-water interface in pulsed power devices flash over. (Montoya, R. and Danneskiold, J. Five seconds at F/16, with a broken camera. *Sandia Lab News* (June 7, 2018). https://www.sandia.gov/news/publications/labnews/articles/2018/08-06/Randy.html. (accessed on 07/24/2020).) If so, then the effective skin depth $\lambda \sim 0$ and the deformation are consistent with the frequencies calculated for the MQN. Computationally or experimentally simulating such a process would be extremely difficult, so surface flashover is consistent with relevant experience but has not been confirmed in the field.

Since the field with the second deformation (the remaining 63 m of trench) has not been drained, its slope (10%) is almost the same as it was in 1868. The simplicity of this particular deformation allows us to estimate the minimum mass of the core of this object from the yield strength and the volume. The product of the minimum pressure $P$, which we equate to the measured yield strength 530 ± 120 kN/m$^2$, times the volume change $\Delta V$, is the minimum work required to compress the trench (1.4 m wide, 1.2 m deep, and 100 m long) and is $\sim 10^8$ joules. The energy for this work came purely from gravitational energy as the "globe of fire" descended the slope from the beginning of the trench to its terminus. The corresponding mass is $m \sim 10^8/(gh) \sim 10^6$ kg, where $g$ is the acceleration of gravity in m/s$^2$ and $h$ is the change in distance toward Earth's center in meters and agrees with the mass estimated from the yield strength and trench diameter.

If the induced currents are flowing in a plasma on the surface, the pressure equals the magnetic pressure $P_m$ produced by the magnetic field $B$:

$$P_m = \frac{B^2}{2\mu_0} \tag{22}$$

Since the pressure supports a mass *m* against the acceleration *g* of gravity, $P_m = mg/(\pi r^2)$ for *r* = the half-width of the trench and $\mu_o = 4\pi \times 10^{-7}$ H/m. The value of *B* spatially and temporarily averaged over the effective area $\pi r^2$ is ~ 5.4 T and is consistent with the spatial and temporal maximum value of *B* = 18.5 T from the simulation.

Fitzgerald reports that the diameter of the luminous ball diminished from ~0.6 m diameter at the beginning to ~0.08 m at the end of the event; however, our survey shows that the width and depth of the depressions in the peat were about the same at the beginning and end of the event. Therefore, the core of the object remained unchanged, with diameter <0.08 m, volume of $<3 \times 10^{-4}$ m$^3$, and density of $>10^6/(3 \times 10^{-4})$ or $>3 \times 10^9$ kg/m$^3$—at least 200,000 times the density of normal matter. Matter does not exist with density between $3 \times 10^9$ kg/m$^3$ and nuclear density. Such a large mass density implies nuclear density matter and is consistent with the ~$10^{18}$ kg/m$^3$ mass density of MQNs.

Based on Table A1 in Appendix D (Tables of MQN Interactions with Water and Granite) and on the MQN's motion to the northeast, the impact must have been >125 km southwest of the 1868 observations for impact velocity ~ 250 km/s. That location is deep in the Atlantic Ocean, >80 km from land, and too far away to be heard by Fitzgerald. Therefore, Table A1 in Appendix D (Tables of MQN Interactions with Water and Granite) applies and indicates the impact was necessarily >200 km from land and is unlikely to be found.

Uncertainties in the current distribution and in applying the quasi-static and uncompressed yield strength to such a dynamic process leads us to assign +/- an order of magnitude error bar to the mass estimate. The resulting $10^{6+/-1}$ kg mass corresponds to the maximum mass in the mass distributions [4] for $B_o = 1.65 \times 10^{12}$ T +/- 21%. For comparison, the magnetic moments and mass densities of protons and neutrons, which are also baryons, correspond to magnetic fields $B_o = 1.5 \times 10^{12}$ T and $2.5 \times 10^{12}$ T, respectively, in reasonable agreement with our value for MQNs. The smaller range is a considerable improvement over Tatsumi's $10^{12+/-1}$ T estimate and permits the design of a systematic test of the MQN dark-matter hypothesis.

## 4. Discussion

The principal result of this paper is reducing uncertainty in the key parameter of the MQN theory to $B_o = 1.65 \times 10^{12}$ T +/- 21%. The result depends on 1) Fitzgerald having accurately reported what he observed, 2) the event being caused by a nearly tangential MQN impact as we have calculated, and 3) the absence of a more likely explanation.

*4.1. Fitzgerald's Accuracy*

We have done what due diligence is possible on Fitzgerald's qualifications as a reliable observer. The County records indicate he was the assistant surveyor for County Donegal during this period. That was a responsible position and is consistent with the detailed report to the Royal Society. We also know that the Royal Society had sufficient confidence in his report to have the president of the Society read it into the proceedings. We have no reason to doubt his integrity.

*4.2. Consistency with MQN Impact*

The details allowed us to find all the secondary observations, i.e., the reported deformations in the peat bog. Carbon 14 analysis, the ordnance survey map of 1860 (prior to the reported event), and the quantitative agreement between our measurements and Fitzgerald's reported measurements, all support consistency with a nearly tangential MQN impact, as we have calculated.

*4.3. Alternative Explanations*

We have attempted to identify other possible causes of the secondary observations. Surface features produced by water erosion of peat and settling under gravity are well documented in the geomorphological literature. Sinuous water-cut gullies dissect peat bogs into complex patterns of haggs (residual masses) and groughs (the gullies) [48]. Mass movement processes involving the flow or slide of water-saturated peat cause major disruptions to bog surfaces and can extend for several hundreds of meters downslope [49]. The features described by Fitzgerald are unlike any of the peat erosional features previously reported and cannot be ascribed to conventional geomorphological processes.

One of us (Professor Peter Wilson) is a geomorphologist specializing in peat bogs. He has investigated the four structures Fitzgerald's eyewitness account connected to the event and concludes that the literature on the geomorphology of peat bogs and his forty years of field work in peat bogs have not suggested any other possible causes for three of the four structures. We conclude that the fourth (the cave) was too close to the stream to preclude its being formed by water flow.

Neutron-stars have the right mass density. However, the gravitational force that holds them together is too weak to sustain the required $\sim 10^6$ kg mass. In addition, pulsar magnetic fields [50] are two orders of magnitude less than those of magnetars [51], upon which the MQN model has been constructed.

Primordial Mini-Black Holes (MBHs) [52] have been proposed to explain luminous and levitating events attributed to anomalous non-weather-related ball lightning [53]. Although an MBH does not have a magnetic field that could levitate it, the net shape of the event horizon from the combined gravitational fields of Earth and a black hole can, in principle, direct evaporating particles downward to provide thrust and levitate the mass [53]. However, the lifetime of a $10^6$ kg MBH is not consistent with the 20-minute event reported by Fitzgerald, and the explosion equivalent to 10 million one-megaton hydrogen bombs characteristic of the final evaporation [52] of an MBH was not observed.

We have also sought alternative explanations from others. Given the eyewitness account and given that our simulations show the MQN hypothesis is consistent with Fitzgerald's eyewitness report, the question becomes can something else reside above and in the peat for 20 minutes, leave meter-scale (depth and width) structures in the peat, and have a glowing light associated with it. We have presented these results to about a thousand people in university colloquia and in contributed and invited talks at meetings of the American Physical Society and Institute of Electrical and Electronic Engineers in the USA, UK, Russia, and India. None of the audience members have suggested an alternative explanation.

*4.4. Limitations to the Evidence*

Even though the three criteria identified in the first paragraph of the Discussion are arguably satisfied, the primary event is the "globe of fire" itself and was seen only by Fitzgerald. The Fitzgerald event is not quite singular. Two other similar primary events qualitatively consistent (i.e., meter-scale, luminous, long-lasting, quasi-spherical, and rotating, as observed by the angular momentum imparted to the surrounding water) with nearly tangential MQN impacts were reported by Soviet Navy Captains at sea in 1962 and 1966 [39]. However, the Fitzgerald event is the only one contemporaneously published in the scientific journal of its day and the only one with detailed secondary observables that can be verified by anyone after the event.

We conclude that the primary event is very rare, is not reproducible, was not recorded by multiple observers, and cannot be quantitatively validated after the fact by anyone else. Consequently, the evidence does not meet today's standard for a discovery.

*4.5. Significance*

However, we also conclude that we can tentatively accept his report and use it to narrow the uncertainty in $B_o$ since it is consistent with an MQN event, and a more likely explanation has not been found. The resulting uncertainty in $B_o$ can be used to design an experiment to systematically test the MQN hypothesis within the constrained range of $B_o$. If MQNs are found as predicted, then the acceptance of the Fitzgerald event as an MQN event will have been validated. If nothing is found, then the experiment will have been another null experiment placing a limit of the mass distribution of MQNs characterized by the tested values of $B_0$, just as all the single-mass quark-nugget experiments to date have been null experiments that placed a flux limit on that single mass.

The results have one additional consequence. Predicted event rates in Figure 2 are so small that even one observed event is consistent with the major portion of dark matter being composed of MQNs. Observing more than one per century suggests a substantial enhancement factor for dark-matter density inside the solar system compared to that of interstellar space. Nuclear-density MQNs are indestructible and can survive passage through the solar photosphere. The magnetopause interaction [28] with matter in the solar photosphere and subsequent deflection by the combined gravity of the planets offer the possibility of enhancing the flux of MQN dark matter within the solar system. Computing an accurate enhancement factor for MQN impacts on Earth requires detailed Monte Carlo simulations beyond the scope of this paper and strongly depends on $B_0$, the radial profile of mass density in the solar photosphere, the velocity distribution of dark matter in interstellar space, and scattering of MQNs by planetary gravity.

In contrast to other candidates for dark matter, MQNs are baryons and, therefore, are consistent [4] with the Standard Model for particles and fields. Well known physics can guide additional experiments and observations [36] to test the MQN hypothesis for dark matter, invent ways for collecting useful MQNs, and develop applications for an indestructible source of $\sim 10^{12}$ tesla magnetic fields.

**5. Conclusions**

The two orders of magnitude uncertainty in Tatsumi's estimate for $B_0$ precludes the practical design of systematic experiments to detect MQNs through their predicted interaction with matter. In this paper, we theoretically examined the signature of a new class of episodic events consistent with a unique signature of MQNs and reported the results of field investigations of one published event consistent with that signature. Tentatively accepting that the event was indeed caused by MQNs constrains the most likely values of $B_0$ to $1.65 \times 10^{12}$ T +/- 21%, which can be used to design a systematic test of the MQN dark-matter hypothesis.

**Author Contributions:** Conceptualization, J.P. Van D. and A.P. Van D.; data curation, J.P. Van D. and N. McG.; formal analysis, J.P. Van D.; Funding acquisition, J.P. Van D.; investigation, J.P. Van D., A.P. Van D., P. W., B.F. H. and N. McG.; methodology, J.P. Van D., A.P. Van D., P. W. and N. McG.; project administration, J.P. Van D.; resources, J.P. Van D.; software, J.P. Van D.; supervision, J.P. Van D.; validation, J.P. Van D., A.P. Van D. and B.F. H.; visualization, J.P. Van D.; writing—original draft, J.P. Van D.; writing— review and editing, J.P. Van D., A.P. Van D., P. W., B.F. H. and N. McG.

All authors have read and agreed to the published version of the manuscript.

**Funding:** This research received no external funding.

**Institutional Review Board Statement:** Not applicable

**Informed Consent Statement:** Not applicable

**Data Availability Statement:** All final analyzed data generated during this study are included in this published article with its supplements.

**Acknowledgments:** We gratefully acknowledge S.V. Greene for first suggesting that quark nuggets might explain the geophysical evidence, and W. F. Brinkman for especially helpful suggestions during this work; Peter van Doorn of the Tornado and Storm Research Organization, UK, for introducing us to the 1868 event; Mr. Josie "The Post" Duddy of County Donegal, Ireland, for showing us the 1985 event; Jesse Rosen for editing the manuscript. We appreciate Mr. and Mrs. William E. Gallagher of Meenawilligan, Kevin Rose, and David Daniels of Booze Allen Hamilton, Arlington, VA, for their assistance in the field work, and Archivists Neve Brennan and Ciara Joyce for locating survey documents from the 1860s.

**Conflicts of Interest:** The authors declare no conflict of interest.

**Appendix A: Quark-Nugget Research Summary**

This summary is an update of one published by us in the open-source article, reference [4]. Therefore, it contains similar information and is included for convenience.

Macroscopic quark nuggets are theoretically predicted objects composed of up, down, and strange quarks in essentially equal numbers. They are also called strangelets [8], nuclearites [9], AQNs [10], and slets [11] and are a subset of Macros [5], a more general term for massive dark matter.

In 1971, Bodmer [12] suggested that a collection of up, down, and strange quarks should be stable. Witten [7] and Farhi and Jaffe [8] showed that quark nuggets should be in the ultra-dense, color-flavor-locked (CFL) phase of quark matter and should be a stable candidate for dark matter. Steiner et al. [13] showed that the ground state of the CFL phase is color neutral and that color neutrality forces electric charge neutrality, which minimizes electromagnetic emissions. However, Xia et al. [11] found that quark depletion causes the ratio $Q/A$ of electric charge $Q$ to baryon number $A$ to be non-zero and varying as $Q/A$ ~0.32 $A^{-1/3}$ for $3 < A < 10^5$. In addition to this core charge, they find that there is a large surface charge and a neutralizing cloud of charge to give a net zero electric charge for sufficiently large $A$. So, quark nuggets with $A \gg 1$ are both dark and very difficult to detect with astrophysical observations.

Bodmer, Witten, and Xia et al. also showed that quark-nugget density should be somewhat larger than the density of nuclei, and their mass can be very large, even as large as the mass of a star. Large quark nuggets are predicted to be stable [7,12-15] with mass between $10^{-8}$ and $10^{20}$ kg within a plausible but uncertain range of assumed parameters of quantum chromodynamics (QCD) and the MIT bag model with its inherent limitations [16].

Although Witten assumed that a first-order phase transition formed quark nuggets, Aoki et al. [17] showed that the finite-temperature QCD transition that formed quark nuggets in the hot early Universe was very likely an analytic crossover, involving a rapid change as the temperature varied but not a real phase transition. Recent simulations by T. Bhattacharya et al. [18] support the crossover process.

A combination of quark nuggets and anti-quark nuggets has also been proposed within constraints imposed by terrestrial observations of the neutrino flux [19]. Zhitnitsky [9] proposed that the collapse of an axion domain-wall network generated Axion Quark Nuggets (AQNs) of both quark and anti-quark varieties. The model relies on the hypothetical axion particle beyond the Standard Model, appears to explain a wide variety of longstanding problems, and leads to AQNs with a narrow mass distribution at ~10 kg [20]. Atreya et al. [21] also found that CP-violating quark and anti-quark scatterings from moving Z(3) domain walls should form quark and anti-quark nuggets, regardless of the order of the quark-hadron phase transition.

Experiments by A. Bazavov et al. [22], at the Relativistic Heavy Ion Collider (RHIC), have provided the first indirect evidence of strange baryonic matter. Additional experiments at RHIC may determine whether the process is a first order phase transition or crossover process. In either case, quark nuggets could have theoretically formed in the early Universe.

In 2001, Wandelt et al. [23] showed that quark nuggets meet the theoretical requirements for dark matter and are not excluded by observations when the stopping power for quark nuggets in the materials covering a detector is properly considered and when the average mass is >$10^5$ GeV (~2 × $10^{-22}$ kg). In 2014, Tulin [24] surveyed additional simulations of increasing sophistication and updated the results of Wandelt, *et al*. The combined results help establish the allowed range and velocity dependence of the strength parameter and strengthen the case for quark nuggets. In 2015, Burdin, *et al.* [25] examined all non-accelerator candidates for stable dark matter and also concluded that quark nuggets meet the requirements for dark matter and have not been excluded experimentally. Jacobs, Starkman, and Lynn [5] found that combined Earth-based, astrophysical, and cosmological observations still allow quark nuggets of mass 0.055 to $10^{14}$ kg and 2 × $10^{17}$ to 4 × $10^{21}$ kg to contribute substantially to dark matter. The large mass means the number per unit volume of space is small, so detecting them requires a very large area detector.

These studies did not consider an intrinsic magnetic field within quark nuggets. However, Xu [26] has shown that the surface magnetic field of quark nuggets from electron ferromagnetism is limited to ~ 2 x $10^7$ T, which is too small for magnetars and MQNs. Tatsumi [27] has shown that, under some special values of the currently unknown QCD coupling constant at the ~ 90 MeV energy scale, a One Gluon Exchange interaction may allow quark nuggets to be ferromagnetic with a surface magnetic field of $10^{12\pm1}$ T. Such a large magnetic field is sufficient for magnetar cores and MQNs. For a quark nugget of radius $r_{QN}$ and a magnetar of radius $r_s$, the magnetic field scales as $(r_{QN}/r_s)^3$. Therefore, the surface magnetic field of a magnetar is smaller than $10^{12}$ T because $r_s > r_{QN}$. Since quark-nugget dark matter is bare, the surface magnetic field of what we wish to detect is $10^{12\pm1}$ T.

Although the cross section for interacting with dense matter is greatly enhanced [28] by the magnetic field, which falls off as radius $r_{QN}^{-3}$, the collision cross section is still many orders of magnitude too small to violate the collision requirements [10,21,22,23] for dark matter.

Chakrabarty [29] showed that the stability of quark nuggets increases with increasing external magnetic field $\leq 10^{16}$ T, so the large self-field described by Tatsumi should enhance their stability. Ping et al. [30] showed that magnetized quark nuggets should be absolutely stable with the newly developed equivparticle model, so the large self-field described by Tatsumi should ensure that quark nuggets with sufficiently large baryon number will not decay by weak interaction.

The large magnetic field also alters MQN interaction with ordinary matter through the greatly enhanced stopping power of the magnetopause around high-velocity MQNs moving through a plasma [28]. Searches [30] for quark nuggets with underground detectors would not be sensitive to highly magnetized quark nuggets, which cannot penetrate the material above the detector. For example, the paper by Gorham and Rotter [19] about constraints on anti-quark-nugget dark matter assumes that limits on the flux of magnetic monopoles from analysis by Price et al. [30] of geologic mica buried under 3 km of rock are also applicable to quark nuggets that can reach the mica.

Porter et al. [32] and Piotrowski et al. [33] reported the absence of sufficiently fast meteor-like objects in the lower atmosphere constrains the flux of quark nuggets (nuclearites) to approximately that required to explain dark matter. Bassan et al. [34] looked for quark nuggets (nuclearites) with gravitational wave detectors and found signals much less than expected for the flux of dark matter.

In summary, experimental or observational evidence of quark nuggets has yet to be found [35] after decades of searching. However, all of these analyses assumed 1) quark nuggets can reach the detector volume because the cross section for momentum transfer is the geometric cross section and 2) all quark nuggets have a single specific mass. In contrast, 1) the MQN magnetopause cross section [28] is many orders of magnitude larger and prevents all but the extremely rare, mostly massive (> 1000 kg) MQNs from being detected and 2) MQNs have a very broad mass distribution [4] which means these experiments do not exclude MQNs.

**Appendix B: COMSOL Simulation of Rotating Magnetized Sphere Interaction with Plastically Deformable Conductor**

The coupled electromagnetic and mechanical interaction between 1) a massive, rotating, strongly magnetized sphere and 2) a nearby conductor was simulated to model the interaction between an MQN and a nearby plastically deformable, electrically conducting medium. The results are provided in the main body of this paper. Additional computational details on the method of this calculation are not of general interest and are given in the following paragraphs.

The overall geometry of the simulation is shown in Figure A1 and consists of a 20 m radius conducting boundary, a cylindrical rotating coordinate system on the same axis as the conducting boundary, a simulated magnetized sphere within the rotating coordinate system, and a 4 × 4 × 2 m slab of simulated peat. The boundary condition for the 20 m radius is **n** × **A** = 0, in which **n** is the unit vector normal to the boundary and **A** is the vector potential, so the magnetic induction vector **B** lies along the boundary surface everywhere. The 20 m radius is sufficiently large to make the force on the peat insensitive to the position of the boundary.

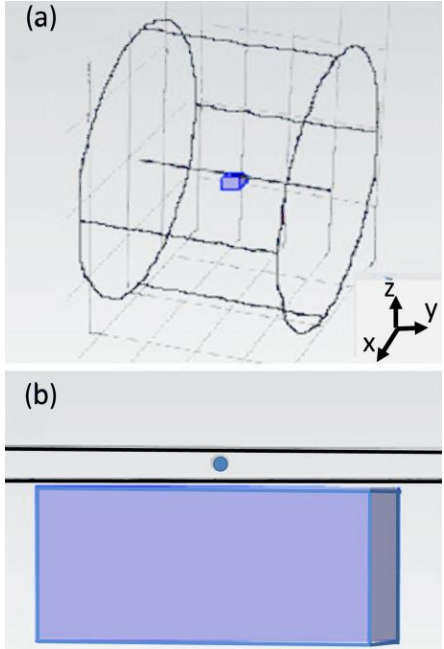

**Figure A1. Geometry of simulation of a rotating quark nugget.** (**a**) The overall geometry of the simulation shows the 20 m radius conducting boundary with the cylindrical rotating coordinate system centered on the axis and the 4 × 4 × 2 m thick slab of simulated peat. (**b**) Close up of the peat slab with its top surface located 0.3 m below the center of the 0.1 m radius, spherical magnet simulating the quark nugget, which is inside the cylindrical 0.2 m radius rotating coordinate system.

The COMSOL [40] Solid Mechanics module solves for the elastic and plastic deformation of the peat under the volume force $\mathbf{F_v} = \mathbf{J} \times \mathbf{B}$, in which $\mathbf{J}$ is the current density in the peat and $\mathbf{B}$ is the magnetic field.

The results are to be compared to observations of an event in peatland. Peat is 80% to 90% water and is not usually studied with solid mechanics models. Therefore, we estimated the mechanical properties of peat for these first calculations based on the properties of water with the measured electrical conductivity and yield stress of peat. Additional work to refine the properties will affect the results somewhat, but the calculated deformation below illuminates the essential behavior. The properties used in this calculation follow:

Electrical conductivity = 22 mS/m for peat. Relative dielectric constant = 80. Relative permeability = 1. Initial density = $10^3$ kg/m$^3$. Yield stress = $5.30 \times 10^5$ Pa. Poisson ratio = 0.4. Young's modulus = $2 \times 10^9$ Pa. Isotropic tangent modulus = $1.1 \times 10^8$ Pa.

Strain is the fractional change, so it is dimensionless. Therefore, the stress versus strain curve is elastic with Young's modulus until a stress of $5.30 \times 10^5$ Pa is exceeded at an elastic strain of $2.0 \times 10^{-4}$. Then the ratio of additional stress to additional strain is the isotropic tangent modulus of $1.1 \times 10^8$ Pa.

The COMSOL Rotating Magnetic Machinery module solves the rotating and non-rotating parts of the problem in their respective coordinate frames and forces continuity of the scalar magnetic potential $V_m$ in the fixed frame. Since the meshes at the interface of the rotating and non-rotating frames are not identical, the calculation interpolates the scalar magnetic potential between the non-conforming meshes. If the magnetic vector potential $\mathbf{A}$ has to be interpolated across the boundary, then current is not conserved. In principle, applying Ampère's law only inside the peat slab avoids this problem. Nevertheless, we varied the radius of the rotating coordinate system to assess the degree to which the numerical interpolation technique affects the results. The choice of the radius of the rotating coordinate system affected the magnetic field at the surface of the peat by approximately ±50%, so this solution is not ideal. However, choosing 0.2 m for the radius of the rotating coordinate system mitigates the problem. This choice gives an air gap of 0.1 m between the interface and the magnetized sphere and between the interface and the surface of the peat.

We used the elastic–plastic deformation module to investigate the deformation of the peat and its dependence on skin depth $\lambda$ with its included frequency $f$, as defined by Equation (21) in the main text. A simulation with the measured conductivity of 22 mS/m and a frequency of $4.5 \times 10^8$ Hz took 1500 seconds of computer time for one 2.2 ns period. Calculating the full deformation at simulation time of 0.03 s would take $6 \times 10^6$ hours of computer time, which is prohibitive. Comparison of simulations at 1, 10, and 100 Hz showed that the force on the peat scales with the product of electrical conductivity $\sigma$ (S/m) and frequency $f$ (Hz), and, therefore, scales with the electromagnetic skin depth $\lambda$.

An intermediate skin depth of 0.05 m, corresponding to $\sigma = 10^7$ S/m and $f = 10$ Hz for $\sigma f = 10^8$ SHz/m, was chosen as the baseline case for these exploratory simulations. The radius of the magnetized sphere was set at 0.1 m and its magnetic induction field was set at 2085 T. The radius of the rotating coordinate system was set at 0.2 m and the front surface of the $4 \times 4 \times 2$ m peat slab was at 0.3 m, as shown in Figure A1.

The simulations showed that the total force on the peat scaled as the square of the magnetic field in the magnetized sphere, as it does in Equation (22). In addition, the magnetic field in the 1868 event, which required a force of $10^7$ N to support the estimated $10^6$ kg mass in Earth's gravitational field, varied as the cube of the radius of the quark nugget, as shown in Figure A2.

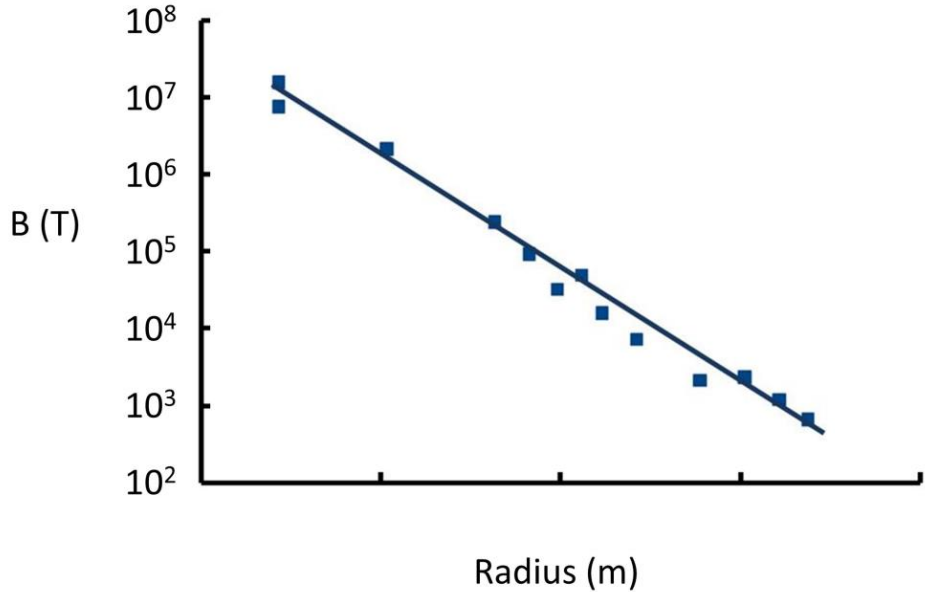

**Figure A2.** Amplitude of the magnetic induction $B$ in the rotating and magnetized sphere required to produce a time-averaged force of $10^7$ N on the peat as a function of the radius of the magnetized sphere. The radius of the magnetized sphere was varied between 5 and 150 mm. The mesh size was too large for the calculation to converge for radii less than 5 mm.

The force required to levitate the sphere for a constant distance from the conducting plane is proportional to the mass of the sphere. The levitating force for the same geometry scales as the square of the surface magnetic field. Therefore, the results are generalized to give the strength of the magnetic field $B_S$ at the surface of the sphere of radius $R_S$ and mass $M_S$:

$$B_S = \frac{C_1 M_S^{0.5}}{R_S^{3 \pm 0.05}} \tag{B1}$$

in which $C_1 = 1.4 \times 10^{-3 \pm 0.3}$ T m³ kg$^{-0.5}$. The coefficient of determination ($R^2$) value is 0.98.

**Appendix C: Field Investigation of M. Fitzgerald's Report to Royal Society**

A "globe of fire" event [38] in County Donegal, Ireland, on August 6, 1868, was reported to the Royal Society by M. Fitzgerald, Assistant Surveyor for County Donegal. His report gave sufficient detail to let us find the deformations he described and investigated the physics that would cause those deformations.

The original publication in the *Proceedings of the Royal Society* [38] describes a series of depressions in peat caused by a "globe of fire". The report is reproduced below, since the only known copy of the original publication is in London, UK, and is only available to researchers approved by the British Library for access.

"On the 6 August 1868, this neighbourhood being free from the dense black clouds that hung over the mountains of Glenswilly and Glendoan, I went up the latter glen to note anything worthy of observation. On arriving at Meenawilligan, the sky was so black over Bintwilly [or Bin Tuile - 'the mountain of floods'], where lightning and thunder were following each other in rapid succession, that I turned homewards. When I reached Folbane, on looking behind, I noticed a globe of fire in the air floating leisurely along in the direction of Church Hill. After passing the crown of the ridge, where I first noticed it, it descended gradually into the valley, keeping all the way about the same distance from the surface of the land, until it reached the stream between Folbane and Derrora, about 300 yards from where I stood. It then struck the land and reappeared in about a minute, drifted along the surface for about 200 yards, and again disappeared into the boggy soil, reappearing about 20 perches (1 perch = 5.03 m) further down the stream; again it moved along the surface, and again sunk, this time into the brow of the stream, which it flew across and finally lodged in the opposite brow, leaving a hole in the peat bank, where it buried itself.

If it had left no marks behind, I confess that, as I had never seen anything of the kind before, I should hesitate to describe its movements, which surprised me much at the time, but the marks which it left behind of its course and power surprised me more.

I at once examined its course, and found a hole about 20 feet square, where it first touched the land, with the pure peat turned out on the lea as if it had been cut out with a huge knife. This was only one minutes work, and, as well as I could judge, it did not occupy fully that time. It next made a drain about 20 perches in length and 4 feet deep, afterwards ploughing up the surface about 1 foot deep, and again tearing away the bank of the stream about 5 perches in length and 5 feet deep, and then hurling the immense mass into the bed of the stream, it flew into the opposite peaty brink. From its first appearance till it buried itself could not have been more than 20 minutes, during which it traveled leisurely, as if floating, with an undulating motion through the air and land over one mile. It appeared at first to be a bright red globular ball of fire, about 2 feet in diameter, but its bulk became rapidly less, particularly after each dip in the soil, so that it appeared not more than 3 inches in diameter when it finally disappeared. The sky overhead was clear at the time but about an hour afterwards it became as dark as midnight."

Fitzgerald's report provided enough information to locate the site of his observations. Since the growth rate of peat in the British Isles during the last few thousand years has generally been in the range of 1-6 cm per century [45] and undisturbed peat readily holds its form, the holes and trenches would still be extant after 137 years. During six separate expeditions to the site in 2004-2006, we found the hole, trench, stream diversion, and cave on privately owned land between 54° 58.321' N, 7° 54.668' W and 54° 58.294' N, 7° 54.576' W. We also made a final visit in 2014 to see how much the site changed over a 10-year period.

Ordnance Survey (OS) maps of 1863 and 1870 were found in the local archive. The 1870 map is a minor revision of the original survey of 1863 rather than a thorough resurvey. A small section of the 1863 map (five years prior to the reported "globe of fire" event of August 6, 1868) is shown in Figure A3. The wavy black lines are the original drawing. Our survey of the area found Fitzgerald's reported features, which are also shown.

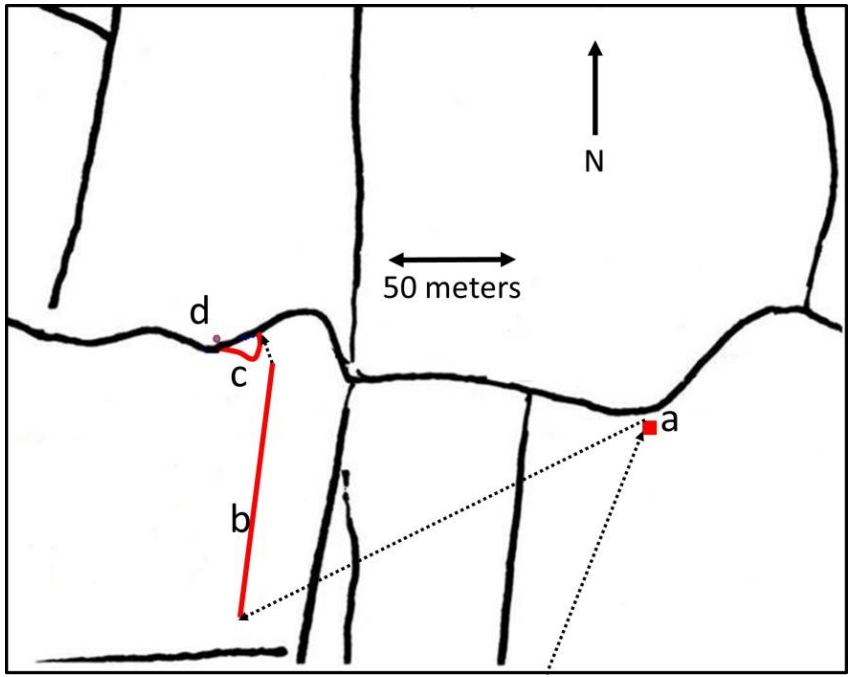

**Figure A3. Extract from the 1863 Ordnance Survey map.** The locations of the "square hole" (a), the most prominent trench (b), triangular channel (c), and cave (d), and the reported path of the "globe of fire" (dotted line) between features are shown. The field with the "20-foot-square" hole (a) has been drained and is lower than the field with the trench (b) and the triangular channel (c). Therefore, we do not know the relative elevation of the hole (a) and the trench (b) in 1868.

The dominant features of the reported depressions are as follows, with letters referring to locations marked in Figure A3:

- Hole (a): ~6.4 m square depression on the course from the crown of the ridge to the south of Meenawilligan towards the town of Churchill.

- Approximately 180 m to the next depression.

- Straight trench (b): ~100 m long, 1.2 m deep, and 1 m wide.

- Unspecified distance to the third depression.

- Curved trench (c): formed when stream bank was "torn away" for 25 m and dumped into the stream.

- Cave (d): a hole in the stream bank directly opposite the end of the "torn away" bank.

Fitzgerald reported that the "globe of fire" first went into the peat bog near the intersection of 1) the line between the "crown of the ridge" and the town of Church Hill and 2) the stream between Derrora and Falabane. At that location, we found a hole of about 6.0 m square with about 0.6 m of open water at 54° 58.294' N and 7° 54.576' W. The hole is located in a marsh at the intersection of a slight west-to-east flow of surface water and two lesser drainage lines coming from the south.

We determined the contours of the hole at 0.3 and 0.5 m below the top of the peat, which are shown in Figure A4. At the 0.5 m depth, the hole is composed of three trenches intersecting at 90-degree angles to form a "square hole". Each trench is approximately 1.2 m wide and 0.8 m to a hard bottom. The length of the south, east, and north sides are, respectively, 2.85, 6.4, and 8.4 m.

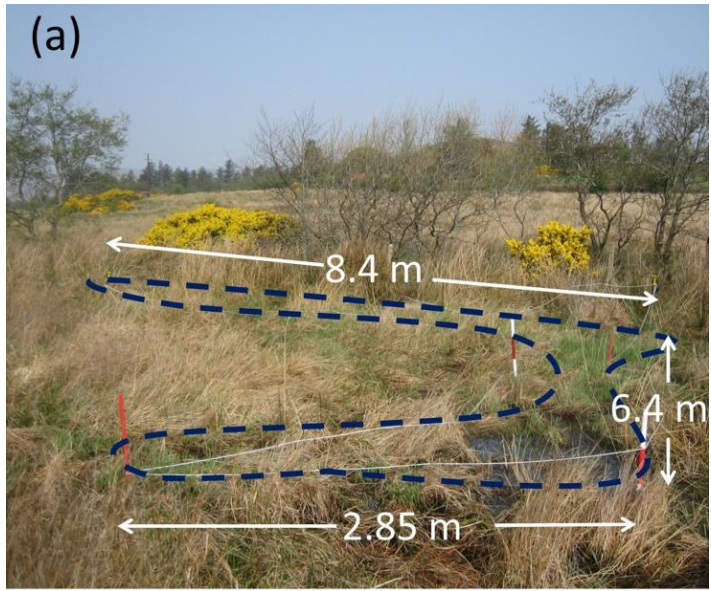

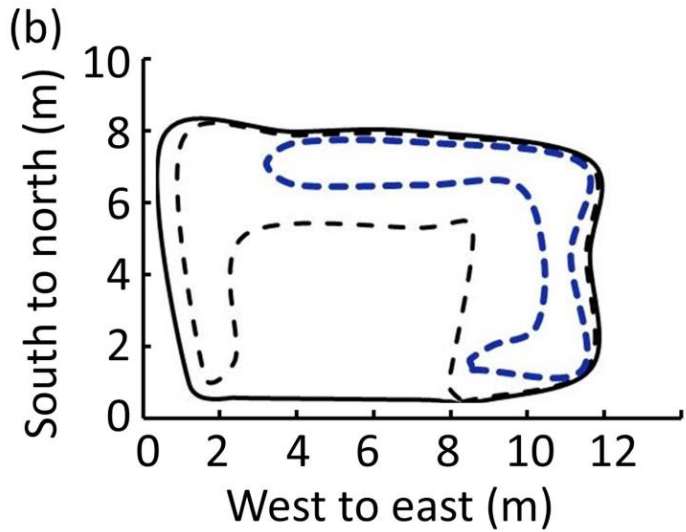

**Figure A4.** The location consistent with Fitzgerald's "20-foot-square hole," at which the ball of light first disappeared into the peat. (a) Photo of the "square hole" with dimensions and 0.5 m deep contour (dashed blue line). (b) Contours at three depths are shown: 0 m (solid black line), 0.3 m (dashed black line), and 0.5 m (dashed blue line). The natural feeder drainage flows approximately from the lower right to the upper left. Orientation to north is approximate.

Fitzgerald reported that the "globe of fire" floated over the bog for about 180 m and then cut a 1.2 m deep trench into the bog for about 100 m. A formerly cultivated field lies approximately 150 m west of the hole and has many north–south-aligned trenches. The seventh trench from the eastern edge of the field is the most prominent and is 63 m long. It lies 0.2 m below the adjacent surface and is approximately 1.2 m wide; these parameters are, respectively, 2.2 and 2.3 standard deviations greater than the mean values of the 26 trenches. The trench is further differentiated from the surrounding terrain by greater penetrability; a ski pole readily penetrates ~0.8 m into the trench but penetrates only ~0.3 m into the surrounding peat with the same force of ~130 N.

Probing the rest of the seventh trench revealed that the firm walls of the trench fall off abruptly (in approximately 0.2 m), indicating a well-defined and deep trench containing low-density peat. However, similar measurements of all 26 trenches show that the penetrability into the peat is not as strong a differentiator between trenches as are depth and width. The probe depth of the seventh trench is only 1.1 standard deviations more than the mean depth of all 26 trenches, and three of the other 25 trenches are deeper. Twelve measurements of the variation in the results give a standard deviation of 0.06 m for this measurement.

This trench is located between 54°58.318′ N, 7°54.651′ W and 54°58.282′ N, 7°54.663′ W and is shown in Figure A5.

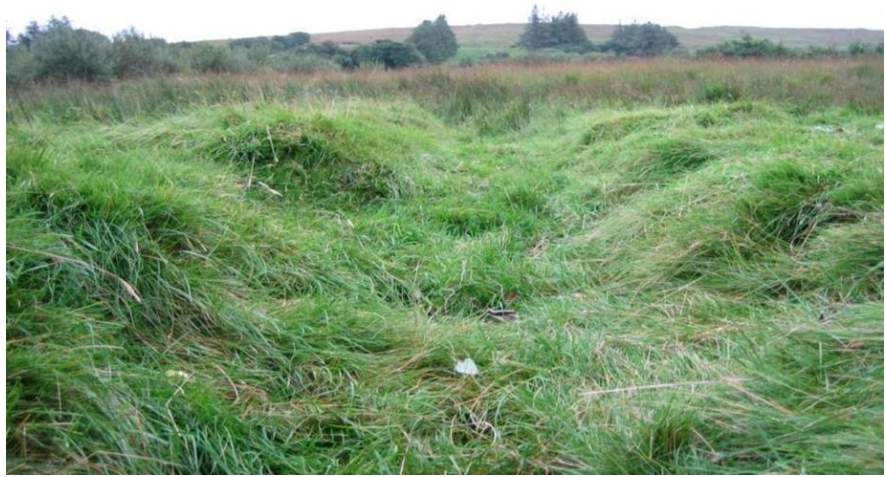

**Figure A5.** Photograph looking along the seventh trench.

Two peat samples from 0.8 ± 0.04 m depth (relative to the top of the adjacent ridges) of the third and seventh trenches and one peat sample at the same depth from the ridge west of the seventh trench were carbon-14 dated by a commercial laboratory. The peat from the most prominent (seventh) trench is 620 ± 60 years old at 85 cm depth and adjacent material is 1330 ± 70 years old at same depth—which is consistent with Fitzgerald's trench, having been filled by erosion with a mixture of peat formed at various times between 0 to 1330 years ago. The next most prominent (third) trench, parallel to the seventh trench, was dated at the same depth as a second control and was found to be 2040 ± 50 years old. This second control is downslope from the most prominent trench and is expected to be older. These data support the uniqueness of the seventh candidate trench. However, measurements in peat are complicated because naturally occurring humic acids can circulate through peat and cause carbon dating to give a later date than isolated material would produce.

The field with these trenches was divided into two parts by the owner in 2000 and separated by a new drainage ditch. The southern portion is used for grazing sheep and is about 30 cm lower than the portion discussed above. The seventh trench ends at the new boundary between the fields at a distance of 63 m from the northern end and 158 m from the hole. If this trench had extended into the newly divided southern portion of the field for another 37 m (to make the ~100 m reported by Fitzgerald), then that end would have been 175 m from the hole and consistent with the ~180 m reported by Fitzgerald. From all evidence above, we can conclude that the seventh trench is the most likely candidate for Fitzgerald's "100-meter long trench".

At its northern (downhill) end, this trench terminates in a mound of peat that prevents it from draining into the stream, and there is no evidence of subsurface piping draining the trench. The mound may have been created when the stream was realigned about 20 years ago, or the trench may have terminated at the mound when it was formed; there is no way to know for certain.

We found a triangular channel to the south of the existing stream and starting at 5 m west of the trench. It is separated from the stream by a mound of mixed mineral and peat debris at 54°58.319' N, 7°54.676' W. The length of the channel is 25 m and its depth is approximately 1.5 m. The landowner said that the water flowed through this channel until the Council redirected the stream in approximately 1989, which agrees with the tree rings observed in one of the trees now growing in the channel bed and with the composition of the mound, which is clearly the material removed from the recently excavated northern channel. The western extremity of this 25 m trench and the stream, as it was recut by the Council, are shown in Figure A6. The photo was taken in 2004 after we cut out the vegetation that had overgrown the trench. In our last trip to the site in 2014, we found the vegetation had again grown throughout the 25 m channel.

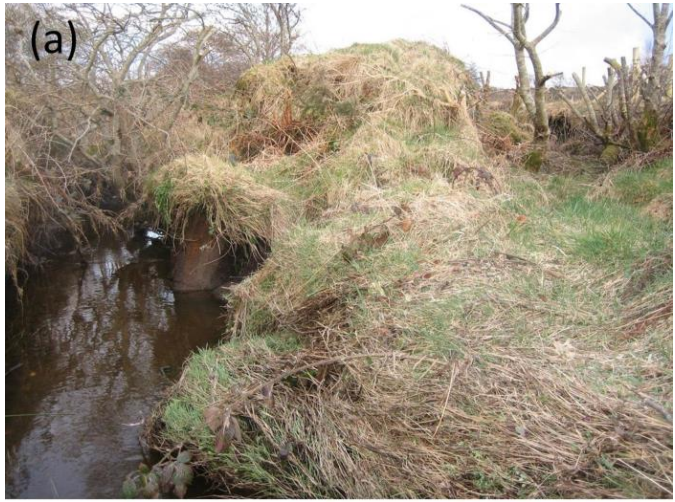

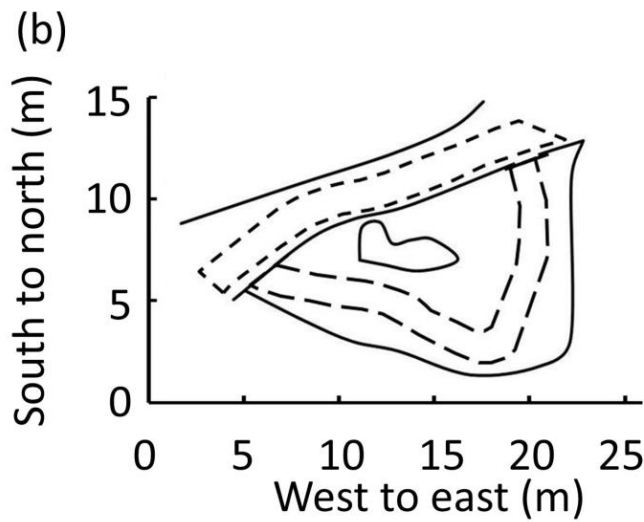

**Figure A6.** Fitzgerald's "25 m long diversion of the stream" and the current path of the stream, which continues to the left and right of the contour map. (a) Photo of the site as seen from the western end; the channel made by the "globe of fire" is on the right and the stream that was recut by the Council is on the left. (b) Contour map of the site constructed from the survey and field notes. Solid lines: surface level. Long dashes: the bottom of the channel at -1.2 ± 0.25 m level. Short dashes: the bottom of the stream, as cut by the County Council in the 1980s, at the -1.9 ± 0.2 m level. Orientation to the north is approximate.

Immediately to the south of the western end of the triangular channel, there is a shallow cave in the stream bank at 54°58.321' N, 7°54.673' W, as shown in Figure A7.

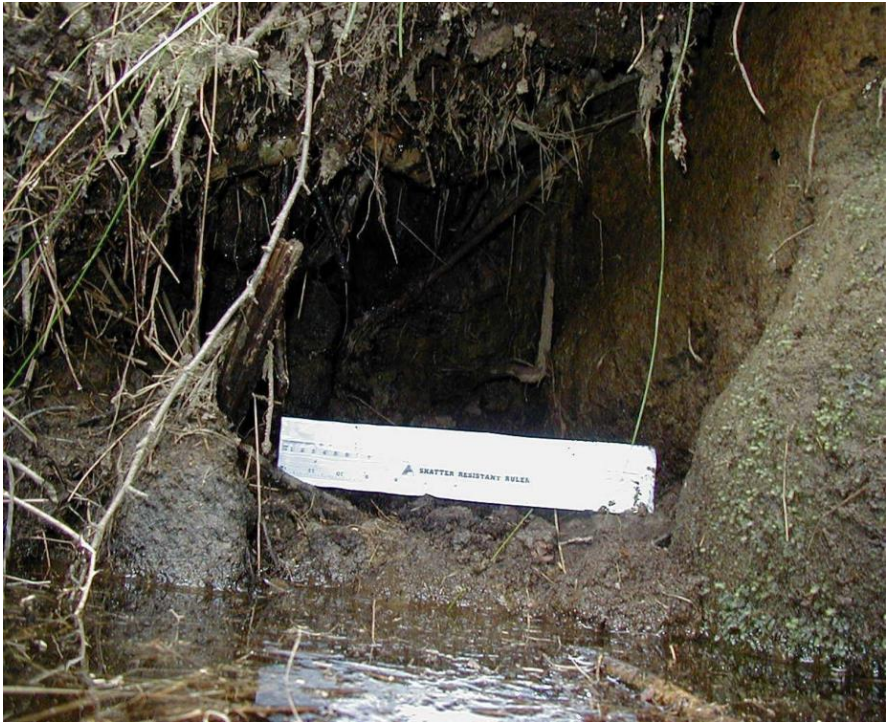

**Figure A7.** Cave at the end of the semi-circular channel.

The depth of the cave is 0.5 m and it is located in the north bank of the stream. The photograph was taken when the water depth was only 10 cm; however, we observed flood debris in stream-bank trees downstream from the cave, indicating that the water rises to at least 1.1 m depth at times. The cave appeared the same in 2014 as it did in 2004.

The electrical resistivity was measured at many places and in several seasons. It was consistently 30 to 60 $\Omega$-m. The compressive yield strength was consistently measured to be 530 ± 120 kN/m² for uncompressed peat, and the strength increased with increasing compression. Radiation measurements were taken. Only background radiation was detected.

Fitzgerald's reported deformations are compared to our findings in Section 3.6 of the main article.

## Appendix D: Tables of MQN Interactions with Water and Granite

**Table A1.** Representative examples are given for MQNs with impact velocity of 250 km/s transiting through water as a function of $B_o$ and MQN mass $m_{qn}$.

| $B_o$ (T) | $1.3\times10^{12}$ | $1.5\times10^{12}$ | $2\times10^{12}$ | $2.5\times10^{12}$ | $3\times10^{12}$ | $3\times10^{12}$ |
|---|---|---|---|---|---|---|
| $m_{qn}$ (kg) | $3.2\times10^{5}$ | $3.2\times10^{5}$ | $3.2\times10^{6}$ | $3.2\times10^{6}$ | $3.2\times10^{6}$ | $3.2\times10^{9}$ |
| $r_{QN}$ (m) for $\rho_{QN} = 10^{18}$ kg/m³ | $4.2\times10^{-5}$ | $4.2\times10^{-5}$ | $9.1\times10^{-5}$ | $9.1\times10^{-5}$ | $9.1\times10^{-5}$ | $9.1\times10^{-4}$ |
| Magnetopause radius $r_m$ (m) | $2.5\times10^{-2}$ | $2.6\times10^{-2}$ | $6.2\times10^{-2}$ | $6.7\times10^{-2}$ | $7.1\times10^{-2}$ | $7.1\times10^{-1}$ |
| Flux for MQN decadal mass (N/y/m²/sr) | $1.4\times10^{-15}$ | $1.4\times10^{-15}$ | $1.5\times10^{-16}$ | $4.3\times10^{-18}$ | $2.7\times10^{-19}$ | $1.3\times10^{-19}$ |
| $x_{max}$ (m) | 241,197 | 219,252 | 389,995 | 336,093 | 297,630 | 2,977,672 |
| $x_{10}$ m/s (m) | 240,914 | 218,995 | 389,539 | 335,700 | 297,282 | 2,974,189 |
| $x_{100}$ m/s (m) | 239,887 | 218,062 | 387,878 | 334,268 | 296,014 | 2,961,506 |
| $\theta_2$ (°) | 88.91690 | 89.01545 | 88.24855 | 88.49068 | 88.66344 | 76.50504 |
| $\theta_{10}$ (°) for $v_{exit} = 10$ m/s | 88.91817 | 89.01660 | 88.25059 | 88.49245 | 88.66500 | 76.52112 |
| $\theta_{100}$ (°) for $v_{exit} = 100$ m/s | 88.92278 | 89.02080 | 88.25806 | 88.49888 | 88.67070 | 76.57968 |
| $t_{exit}$ (s) for $v_{exit} = 10$ m/s | $5.4\times10^{1}$ | $5.0\times10^{1}$ | $8.8\times10^{1}$ | $7.6\times10^{1}$ | $6.7\times10^{1}$ | $6.7\times10^{2}$ |
| $t_{exit}$ (s) for $v_{exit} = 100$ m/s | $2.4\times10^{1}$ | $2.2\times10^{1}$ | $3.9\times10^{1}$ | $3.4\times10^{1}$ | $3.0\times10^{1}$ | $3.0\times10^{2}$ |
| $\delta$ fractional error for $v_{exit} = 10$ m/s | $6.0\times10^{-2}$ | $5.5\times10^{-2}$ | $9.7\times10^{-2}$ | $8.4\times10^{-2}$ | $7.4\times10^{-2}$ | $6.0\times10^{-1}$ |
| $\delta$ fractional error for $v_{exit} = 100$ m/s | $1.2\times10^{-2}$ | $1.1\times10^{-2}$ | $1.9\times10^{-2}$ | $1.7\times10^{-2}$ | $1.5\times10^{-2}$ | $1.5\times10^{-1}$ |
| Cross section for all $v_{exit}$ | $4.6\times10^{10}$ | $3.8\times10^{10}$ | $1.2\times10^{11}$ | $8.9\times10^{10}$ | $7.0\times10^{10}$ | $7.1\times10^{12}$ |
| Cross section for $v_{exit} = 10$ to 100 m/s | $3.9\times10^{8}$ | $3.2\times10^{8}$ | $1.0\times10^{9}$ | $7.5\times10^{8}$ | $5.9\times10^{8}$ | $6.1\times10^{10}$ |
| Total number per year | $3.5\times10^{-4}$ | $2.9\times10^{-4}$ | $9.7\times10^{-5}$ | $2.1\times10^{-6}$ | $1.0\times10^{-7}$ | $5.2\times10^{-6}$ |
| Number per year for 10 to 100 m/s $v_{exit}$ | $3.0\times10^{-6}$ | $2.5\times10^{-6}$ | $8.2\times10^{-7}$ | $1.8\times10^{-8}$ | $8.7\times10^{-10}$ | $4.5\times10^{-8}$ |
| Frequency (MHz) | $7.0\times10^{0}$ | $7.0\times10^{0}$ | $3.2\times10^{0}$ | $3.1\times10^{0}$ | $3.0\times10^{0}$ | $3.1\times10^{-1}$ |
| Rotational energy (J) | $1.4\times10^{5}$ | $1.3\times10^{5}$ | $1.3\times10^{6}$ | $1.2\times10^{6}$ | $1.2\times10^{6}$ | $1.2\times10^{9}$ |
| RF power (MW) | $4.4\times10^{3}$ | $5.6\times10^{3}$ | $4.3\times10^{4}$ | $6.2\times10^{4}$ | $8.3\times10^{4}$ | $8.4\times10^{6}$ |
| RF power (MW) after 1,200 s | $6.5\times10^{0}$ | $5.0\times10^{0}$ | $6.0\times10^{1}$ | $3.9\times10^{1}$ | $2.8\times10^{1}$ | $2.0\times10^{5}$ |

**Table A2.** Representative examples are given for MQNs with impact velocity of 250 km/s transiting through granite as a function of $B_o$ and MQN mass $m_{qn}$.

| $B_o$ (T) | $1.3\times10^{12}$ | $1.5\times10^{12}$ | $2\times10^{12}$ | $2.5\times10^{12}$ | $3\times10^{12}$ | $3\times10^{12}$ |
|---|---|---|---|---|---|---|
| $m_{qn}$ (kg) | $3.2\times10^5$ | $3.2\times10^5$ | $3.2\times10^6$ | $3.2\times10^6$ | $3.2\times10^6$ | $3.2\times10^9$ |
| $r_{QN}$ (m) for $\rho_{QN} = 10^{18}$ kg/m³ | $4.2\times10^{-5}$ | $4.2\times10^{-5}$ | $9.1\times10^{-5}$ | $9.1\times10^{-5}$ | $9.1\times10^{-5}$ | $9.1\times10^{-4}$ |
| Magnetopause radius $r_m$ (m) | $2.2\times10^{-2}$ | $2.3\times10^{-2}$ | $5.4\times10^{-2}$ | $5.8\times10^{-2}$ | $6.2\times10^{-2}$ | $6.2\times10^{-1}$ |
| Flux for MQN decadal mass (N/y/m2/sr) | $1.4\times10^{-15}$ | $1.4\times10^{-15}$ | $1.5\times10^{-16}$ | $4.3\times10^{-18}$ | $2.7\times10^{-19}$ | $1.3\times10^{-19}$ |
| $x_{max}$ (m) | 138,434 | 125,839 | 223,837 | 192,900 | 170,824 | 1,709,027 |
| $x_{10}$ m/s (m) | 138,272 | 125,692 | 223,575 | 192,674 | 170,624 | 1,707,028 |
| $x_{100}$ m/s (m) | 137,683 | 125,156 | 222,622 | 191,852 | 169,897 | 1,699,749 |
| $\theta_2$ (°) | 89.37838 | 89.43494 | 88.99486 | 89.13380 | 89.23293 | 82.30288 |
| $\theta_{10}$ (°) for $v_{exit} = 10$ m/s | 89.37911 | 89.43560 | 88.99604 | 89.13481 | 89.23383 | 82.31194 |
| $\theta_{100}$ (°) for $v_{exit} = 100$ m/s | 89.38176 | 89.43801 | 89.00032 | 89.13850 | 89.23710 | 82.34492 |
| $t_{exit}$ (s) for $v_{exit} = 10$ m/s | $3.1\times10^1$ | $2.8\times10^1$ | $5.1\times10^1$ | $4.4\times10^1$ | $3.9\times10^1$ | $3.9\times10^2$ |
| $t_{exit}$ (s) for $v_{exit} = 100$ m/s | $1.4\times10^1$ | $1.3\times10^1$ | $2.3\times10^1$ | $1.9\times10^1$ | $1.7\times10^1$ | $1.7\times10^2$ |
| $\delta$ fractional error for $v_{exit} = 10$ m/s | $3.5\times10^{-2}$ | $3.1\times10^{-2}$ | $5.6\times10^{-2}$ | $4.8\times10^{-2}$ | $4.3\times10^{-2}$ | $3.9\times10^{-1}$ |
| $\delta$ fractional error for $v_{exit} = 100$ m/s | $6.9\times10^{-3}$ | $6.3\times10^{-3}$ | $1.1\times10^{-2}$ | $9.6\times10^{-3}$ | $8.5\times10^{-3}$ | $8.5\times10^{-2}$ |
| Cross section for all $v_{exit}$ | $1.5\times10^{10}$ | $1.2\times10^{10}$ | $3.9\times10^{10}$ | $2.9\times10^{10}$ | $2.3\times10^{10}$ | $2.3\times10^{12}$ |
| Cross section for $v_{exit} = 10$ to 100 m/s | $1.3\times10^8$ | $1.1\times10^8$ | $3.3\times10^8$ | $2.5\times10^8$ | $1.9\times10^8$ | $2.0\times10^{10}$ |
| Total number per year | $1.2\times10^{-4}$ | $9.7\times10^{-5}$ | $3.2\times10^{-5}$ | $6.9\times10^{-7}$ | $3.4\times10^{-8}$ | $1.7\times10^{-6}$ |
| Number per year for 10 to 100 m/s $v_{exit}$ | $9.9\times10^{-7}$ | $8.2\times10^{-7}$ | $2.7\times10^{-7}$ | $5.9\times10^{-9}$ | $2.9\times10^{-10}$ | $1.4\times10^{-8}$ |
| Frequency (MHz) | $9.0\times10^0$ | $8.9\times10^0$ | $4.0\times10^0$ | $3.9\times10^0$ | $3.9\times10^0$ | $3.9\times10^{-1}$ |
| Rotational energy (J) | $2.2\times10^5$ | $2.2\times10^5$ | $2.1\times10^6$ | $2.0\times10^6$ | $2.0\times10^6$ | $1.9\times10^9$ |
| RF power (MW) | $1.2\times10^4$ | $1.5\times10^4$ | $1.1\times10^5$ | $1.6\times10^5$ | $2.2\times10^5$ | $2.2\times10^7$ |
| RF power (MW) after 1,200 s | $6.7\times10^0$ | $5.1\times10^0$ | $6.1\times10^1$ | $4.0\times10^1$ | $2.8\times10^1$ | $2.3\times10^5$ |